# First-Principle Homogenization Theory for Periodic Metamaterials

Andrea Alù

Department of Electrical and Computer Engineering, The University of Texas at Austin, Austin, TX 78712, USA, alu@mail.utexas.edu

**Abstract:** *We derive from first principles an accurate homogenized description of periodic metamaterials made of magnetodielectric inclusions, highlighting and overcoming relevant limitations of standard homogenization methods. We obtain closed-form expressions for the effective constitutive parameters, pointing out the relevance of inherent spatial dispersion effects, present even in the long-wavelength limit. Our results clarify the limitations of quasi-static homogenization models, restore the physical meaning of homogenized metamaterial parameters and outline the reasons behind magnetoelectric coupling effects that may arise also in the case of center-symmetric inclusions.*

1. Introduction and Motivation

The electromagnetic homogenization of natural and artificial materials has a long-standing tradition [1]-[7] and several theories are available to define macroscopic averaged quantities that represent the effective constitutive parameters of periodic or random collections of molecules or inclusions. The same way in which we define permittivity and permeability of natural materials, by averaging out irrelevant microscopic field fluctuations at the atomic or molecular level, also in the field of artificial materials and mixtures homogenization and mixing rules have been put forward over the years to avoid solving for the complex electromagnetic interaction among a large number of inclusions [7]. Homogenized descriptions of natural and artificial materials can hold only in the long-wavelength regime, i.e., for effective wavelengths and averaged field variations much larger than the material granularity. Within these limits, such descriptions have proven to be accurate and of great advantage for analysis and design purposes.

With the advent of metamaterials [8], i.e., artificial materials with anomalous and exotic electromagnetic response, the necessity of more advanced concepts and improved homogenization models has become evident, since often the topological and/or resonant properties of the inclusions and building blocks do not allow a description of their properties in terms of simple averaging procedures. Often, the exotic metamaterial properties are inherently based on these anomalous features, which cannot be captured by simple homogenization schemes inspired to natural materials. The necessity for improved models has been outlined in several recent papers on the topic [9]-[26], which describe different approaches to the problem.

The simplest homogenization technique consists of retrieving the effective parameters from the scattering properties of a metamaterial sample. The Nicholson-Ross-Weir (NRW) retrieval method postulates the equivalence between a complex metamaterial array and a uniform slab of same thickness with unknown constitutive parameters, usually limited to permittivity and permeability [27]-[29]. This approach, appealing for its simplicity, often provides constitutive parameters with nonphysical frequency dispersion, in particular near the inclusion resonances, yielding complex values of permittivity and permeability that violate basic passivity and causality constraints [30]. The typical presence of "anti-resonant" artifacts in the dispersion of the effective constitutive parameters, wrong sign of their imaginary parts, wrong slope of their real part and inherent dependence of these parameters on the metamaterial boundaries and excitation [31] are all clear signs of the inadequacy of this approach when applied to resonant artificial materials, as more extensively discussed in [32].

In the recent literature, these anomalies have been generically related to strong spatial dispersion, which should be taken into account in more refined homogenization models. In this context, analytical and semi-analytical methods have been put forward to address the homogenization in a more rigorous fashion. Generalized Clausius-Mossotti techniques have been extended to the case of complex inclusions, including bianisotropic effects, possible presence of spatial dispersion and accurate modeling of the inclusion interaction [14]-[22]. As another successful approach, averaging a planar sheet of inclusions and then considering the mutual interaction among parallel layers as a Bloch lattice has also been proposed [9]-[13]. In these schemes too, however, spatial dispersion can often generate artifacts and dependence of the extracted effective parameters on the choice of excitation and boundary conditions, effects that are not easily explained on clear physical grounds.

In order to circumvent these issues, a rigorous approach to the homogenization of periodic arrays of dielectric inclusions [23]-[26] has been put forward based on the Floquet representation used in optical crystals [33], i.e., by introducing a single generalized permittivity tensor that includes all the polarization effects, including artificial magnetism, bianisotropy and higher-order spatial dispersion effects. This technique is limited to dielectric-only periodic metamaterials and the single-permittivity representation may often make challenging to relate weak spatial dispersion effects in the generalized permittivity tensor to artificial magnetic or bianisotropic effects. It would be more desirable to describe these effects in terms of local permeability or chirality parameters [34], whenever possible. In addition, the spatial dispersion properties of the generalized permittivity tensor makes more challenging to apply usual boundary conditions and dispersion relations valid for metamaterials with local properties. In most circumstances, for frequencies well below the inclusion resonances, for which both the background wavelength and the effective guided wavelength are significantly larger than the average array granularity, local constitutive parameters should be sufficient to characterize the metamaterial response, and spatial dispersion effects may be negligible [35].

In this paper, we develop from first-principles a new homogenization theory that combines the advantages of currently available homogenization schemes, without their drawbacks and limitations: we extend the rigorous approach of Floquet-based theories [23],[33] to magnetodielectric metamaterials and, proposing a different averaging scheme, we combine it with the generality of retrieval techniques and with the convenience of local homogenization schemes, extending their validity and applicability to resonant metamaterials. We develop a self-consistent Floquet homogenization theory for metamaterial arrays formed by arbitrary electric and magnetic inclusions, which can rigorously take into account the complex wave interaction among inclusions, and which does not depend on the form of excitation, converging to a local model in the long-wavelength limit.

Our analysis clarifies limits and approximations of other homogenization techniques, highlighting the reasons and physical mechanisms behind artifacts and nonphysical dispersion of homogenized parameters and demonstrating that a rigorous analysis of the array coupling inherently requires considering frequency and spatial dispersion effects at the lattice level, even when higher-order multipolar interaction or bianisotropic effects within the unit cell are negligible. These effects modify the usual form of effective constitutive models and are associated with a direct manifestation of the finite phase velocity with respect to the array period, particularly relevant in the case of more densely packed, and possibly resonant, metamaterials.

Our findings establish the foundations of a new, physically meaningful description of a wide class of metamaterials, valid even when the density of inclusions is not small and classic homogenization models, like Clausius-Mossotti relations [7], lose their accuracy even in the long wavelength regime.

The paper is organized as follows: in Section 2, a Floquet-based homogenization approach on the model of [23],[33] is generalized to the presence of electric and magnetic materials and arbitrary sources. This generalization is particularly important, since it will allow defining, for the first time to our knowledge, metamaterial constitutive parameters that inherently do not depend on the form and nature of excitation. In addition, we introduce a Taylor expansion of the polarization and magnetization currents to derive a self-consistent definition of averaged fields, which is able to extract weak spatial dispersion effects and allows an averaged local description in the long-wavelength limit. In Section 3, closed-form expressions for the new *effective* constitutive parameters are derived, highlighting the mentioned advantages of independence on the applied sources and on the wave vector in the long-wavelength limit. In Section 4, this general theory is applied to the special circumstance of absence of impressed sources, showing that in this case the general constitutive relations may be written in terms of *equivalent* constitutive parameters, which coincide with those obtainable using simple retrieval procedures. This alternative model is shown to hide inherent spatial dispersion effects and nonphysical features, and its use should be limited to the solution of practical scattering problems in absence of impressed sources. This discussion will highlight the limitations and inherent approximations of other homogenization schemes and will provide physical insights into the more rigorous averaged description of metamaterials introduced here. Sections 5 and 6 analyze the homogenization model in the long-wavelength and in the resonant limits, regions of special interest for metamaterial applications. Finally, Section 7 validates our theory with numerical examples and further discussions.

## 2. General Homogenization Theory for Periodic Magnetodielectric Metamaterials

In this section, as the first objective of this paper, we develop from first-principles a general homogenization theory for periodic arrays of arbitrarily shaped dielectric, magnetic and/or conducting inclusions, extending the rigorous Floquet approach commonly used in optical crystals [33] and dielectric metamaterials [23] to arbitrary inclusions and arbitrary form of excitation. We also define a new averaging procedure by using a Taylor expansion of the microscopic field variations, which will allow defining constitutive parameters that have local properties in the long-wavelength limit. For simplicity of notation, we assume here a cubic lattice with period $d$, but extension to arbitrary lattices may also be envisioned.

The most general description of a periodic array in its linear operation may be developed, without loss of generality, in the Fourier domain [33]. Our goal is to derive the general form of macroscopic constitutive relations for any arbitrary pair $(\omega, \boldsymbol{\beta})$, relating spatially averaged field quantities that vary as $e^{i\boldsymbol{\beta}\cdot\mathbf{r}}e^{-i\omega t}$. Only a limited set of eigenvectors $\boldsymbol{\beta}$ are supported by the array at a given frequency $\omega$ in the absence of impressed sources. These correspond to the eigenmodes of the system, which are usually the focus of homogenization theories and will be analyzed in detail in Section 4. However, an average description of the array as a bulk material should not depend on the possible presence of impressed embedded sources or on the relative amplitude of electric and magnetic fields at a specific point in space. Therefore, we assume here the presence of impressed sources with arbitrary $e^{i\boldsymbol{\beta}\cdot\mathbf{r}}e^{-i\omega t}$ plane-wave like dependence, uniformly distributed all over the array. This ensures an averaged space-time distribution of the induced fields with the same $e^{i\boldsymbol{\beta}\cdot\mathbf{r}}e^{-i\omega t}$ dependence, in which the variables $\omega, \boldsymbol{\beta}$ are independent of each other. In practice, it may be challenging to realize a distribution of uniformly impressed sources with plane-wave dependence within a metamaterial array, so this excitation should be interpreted as a test excitation to isolate the metamaterial response in the Fourier domain, or as a specific Fourier component of embedded sources with localized space-time distribution. We will specialize these results to eigenmodal propagation (source-free scenario) in Section 4.

In the most general case, the *microscopic* [37] field distribution at any point in the array satisfies

$$\nabla \times \mathbf{E}(\mathbf{r}) = i\omega\mu_0 \mathbf{H}(\mathbf{r}) + i\omega \mathbf{M}(\mathbf{r}) - \mathbf{K}_{ext} e^{i\boldsymbol{\beta}\cdot\mathbf{r}}$$
$$\nabla \times \mathbf{H}(\mathbf{r}) = -i\omega\varepsilon_0 \mathbf{E}(\mathbf{r}) - i\omega \mathbf{P}(\mathbf{r}) + \mathbf{J}_{ext} e^{i\boldsymbol{\beta}\cdot\mathbf{r}}, \quad (1)$$

where $\mathbf{E}(\mathbf{r})$, $\mathbf{H}(\mathbf{r})$ are the local electric and magnetic fields, $\mathbf{P}(\mathbf{r})$ is the local polarization vector, $\mathbf{M}(\mathbf{r})$ is the local magnetization vector, $\mathbf{J}_{ext}$ and $\mathbf{K}_{ext}$ are complex vectors of independently impressed distributed electric and magnetic current density sources with explicit plane-wave dependence $e^{i\boldsymbol{\beta}\cdot\mathbf{r}}$, and $\varepsilon_0$, $\mu_0$ are the background permittivity and permeability, respectively. Due to the linearity of the problem, we have suppressed in (1) a common $e^{-i\omega t}$ time dependence. In the presence of electric or magnetic conductors, the induced current densities $\mathbf{J}_{ind}$, $\mathbf{K}_{ind}$ are implicitly embedded in $\mathbf{P}(\mathbf{r}) = i\mathbf{J}_{ind}(\mathbf{r})/\omega$ and $\mathbf{M}(\mathbf{r}) = i\mathbf{K}_{ind}(\mathbf{r})/\omega$ in Eq. (1). The distributed impressed source distributions may also be seen as sustaining impressed fields with the same $e^{i\boldsymbol{\beta}\cdot\mathbf{r}}e^{-i\omega t}$ plane-wave dependence and complex amplitudes satisfying

$$i\boldsymbol{\beta} \times \mathbf{E}_{ext} = i\omega\mu_0 \mathbf{H}_{ext} - \mathbf{K}_{ext}$$
$$i\boldsymbol{\beta} \times \mathbf{H}_{ext} = -i\omega\varepsilon_0 \mathbf{E}_{ext} + \mathbf{J}_{ext}. \quad (2)$$

Notice that the arbitrary choice of $\mathbf{J}_{ext}$ and $\mathbf{K}_{ext}$ in (1) implies that the complex amplitudes of impressed fields $\mathbf{E}_{ext}$ and $\mathbf{H}_{ext}$ are independent of each other. This will be very important to ensure the general validity of the effective homogenization model proposed here, as discussed below.

Due to the periodicity of the crystal, we may write Eq. (1) in the $e^{i\boldsymbol{\beta}\cdot\mathbf{r}}e^{-i\omega t}$ Fourier domain

$$i\boldsymbol{\beta}\times\bar{\mathbf{E}} = i\omega\mu_0\bar{\mathbf{H}} + i\omega\bar{\mathbf{M}} - \mathbf{K}_{ext}$$
$$i\boldsymbol{\beta}\times\bar{\mathbf{H}} = -i\omega\varepsilon_0\bar{\mathbf{E}} - i\omega\bar{\mathbf{P}} + \mathbf{J}_{ext},$$
(3)

where the bar denotes the averaging operation $\bar{\mathbf{E}} = \frac{1}{d^3}\int_V \mathbf{E}(\mathbf{r})e^{-i\boldsymbol{\beta}\cdot\mathbf{r}}d\mathbf{r}$, and similarly for all the other vectors in Eq. (3). This averaging procedure, consistent with [33],[23], filters out the dominant contribution to the local field $\mathbf{E}(\mathbf{r})$, varying as $\bar{\mathbf{E}}e^{i\boldsymbol{\beta}\cdot\mathbf{r}}$, of interest for a macroscopic homogenized description of the array. Eq. (3) relates the complex amplitudes of the spatially averaged *macroscopic* [37] field quantities, which all vary with an implicit $e^{i\boldsymbol{\beta}\cdot\mathbf{r}}e^{-i\omega t}$ space-time dependence due to the chosen form of impressed excitation and the linearity of the problem. Inspecting Eq. (3), one may be tempted to define spatially averaged displacement vectors as $\bar{\mathbf{B}} = \mu_0\bar{\mathbf{H}} + \bar{\mathbf{M}}$ and $\bar{\mathbf{D}} = \varepsilon_0\bar{\mathbf{E}} + \bar{\mathbf{P}}$, and the associated constitutive relations $\bar{\mathbf{B}} = \underline{\boldsymbol{\mu}}_g \cdot \bar{\mathbf{H}}$, $\bar{\mathbf{D}} = \underline{\boldsymbol{\varepsilon}}_g \cdot \bar{\mathbf{E}}$, which would generalize the metamaterial homogenization approach used in [33],[23] to the case of magnetodielectric materials. However, this macroscopic description has several shortcomings: the permittivity $\underline{\boldsymbol{\varepsilon}}_g$ and permeability $\underline{\boldsymbol{\mu}}_g$ respectively coincide with $\varepsilon_0$ and $\mu_0$ when the inclusions are formed, at the microscopic level, by purely magnetic ($\mathbf{P}(\mathbf{r}) = \bar{\mathbf{P}} = \mathbf{0}$) or dielectric ($\mathbf{M}(\mathbf{r}) = \bar{\mathbf{M}} = \mathbf{0}$) materials, respectively. This implies that artificial magnetic or polarization effects, stemming from the rotation of electric or magnetic polarization respectively, remain hidden as spatial dispersion effects in the permittivity $\underline{\boldsymbol{\varepsilon}}_g$ or permeability $\underline{\boldsymbol{\mu}}_g$ tensors. In particular, $\underline{\boldsymbol{\varepsilon}}_g$ coincides with the generalized permittivity defined in [23] in the case of dielectric-only or conducting inclusions. This description, therefore, cannot converge to a local constitutive model in the long-wavelength limit in the presence of common artificial magnetic or dielectric effects. For instance, in the case of a metamaterial formed by conducting split-ring resonators [25], this model would predict $\mathbf{M}(\mathbf{r}) = \overline{\mathbf{M}} = \mathbf{0}$, $\underline{\boldsymbol{\mu}}_g = \mu_0 \underline{\mathbf{I}}$ (with $\underline{\mathbf{I}}$ being the identity matrix), despite the evident presence of magnetic effects, which remain hidden in the weak spatial dispersion of $\underline{\boldsymbol{\varepsilon}}_g$. We propose in the next subsection a different averaging scheme that takes these effects into account and provides a homogenized description converging to a local model in the long-wavelength limit.

*a) Multipolar expansion*

In order to overcome the issue outlined above, we assume that the unit cell is sufficiently smaller than the wavelength of operation to ensure that the induced microscopic polarization and magnetization vectors slowly vary within each unit cell, as usual in metamaterials. In such circumstances, it is possible to expand $\overline{\mathbf{P}}$ in a Taylor series around the origin of each unit cell, to obtain [38]

$$\overline{\mathbf{P}} = \frac{1}{d^3} \int_V \mathbf{P}(\mathbf{r}) e^{-i\boldsymbol{\beta}\cdot\mathbf{r}} d\mathbf{r} =$$

$$= \frac{1}{d^3} \left[ \begin{array}{l} \int_V \mathbf{P}(\mathbf{r}) d\mathbf{r} + i\boldsymbol{\beta} \times \int_V \frac{\mathbf{r} \times \mathbf{P}(\mathbf{r})}{2} d\mathbf{r} - \frac{i\boldsymbol{\beta}}{2} \cdot \int_V \left[ \mathbf{r} \mathbf{P}(\mathbf{r}) + \mathbf{P}(\mathbf{r})\mathbf{r} \right] d\mathbf{r} + \\ -\boldsymbol{\beta} \times \int_V \frac{\mathbf{r} \times \mathbf{P}(\mathbf{r}) \times \mathbf{r}}{6} d\mathbf{r} \times \boldsymbol{\beta} + \frac{\boldsymbol{\beta}}{2} \times \left[ \boldsymbol{\beta} \cdot \int_V \frac{\left[ \mathbf{r} \times \mathbf{P}(\mathbf{r}) \right] \mathbf{r} + \mathbf{r} \left[ \mathbf{r} \times \mathbf{P}(\mathbf{r}) \right]}{3} d\mathbf{r} \right] + ... \end{array} \right] =,\quad (4)$$

$$= \mathbf{P}_E - \frac{\boldsymbol{\beta} \times \overline{\mathbf{M}}_E}{\omega \mu_0} - \frac{i\boldsymbol{\beta}}{2} \cdot \underline{\mathbf{Q}}_E^e + \frac{\boldsymbol{\beta}}{i\omega} \times \left[ \overline{\mathbf{P}}_E' \times \boldsymbol{\beta} - \frac{\boldsymbol{\beta} \cdot \underline{\mathbf{Q}}_E^m}{2} \right] + ...$$

where

$$\begin{aligned}
\mathbf{P}_E &= \frac{1}{d^3}\int_V \mathbf{P}(\mathbf{r})\,d\mathbf{r} \\
\mathbf{M}_E &= -\frac{i\omega\mu_0}{d^3}\int_V \frac{\mathbf{r}\times\mathbf{P}(\mathbf{r})}{2}\,d\mathbf{r} \\
\underline{\mathbf{Q}}_E^e &= \frac{1}{d^3}\int_V \big[\mathbf{r}\mathbf{P}(\mathbf{r})+\mathbf{P}(\mathbf{r})\mathbf{r}\big]\,d\mathbf{r} \\
\mathbf{P}'_E &= -\frac{i\omega}{d^3}\int_V \frac{\mathbf{r}\times\mathbf{P}(\mathbf{r})\times\mathbf{r}}{6}\,d\mathbf{r} \\
\underline{\mathbf{Q}}_E^m &= -\frac{i\omega}{d^3}\int_V \frac{[\mathbf{r}\times\mathbf{P}(\mathbf{r})]\mathbf{r}+\mathbf{r}[\mathbf{r}\times\mathbf{P}(\mathbf{r})]}{3}\,d\mathbf{r}
\end{aligned} \qquad (5)$$

represent the first electric and magnetic multipole moments associated with the induced electric polarization distribution $\mathbf{P}(\mathbf{r})$. In particular, $\mathbf{P}_E$, $\mathbf{M}_E$ are the fist-order contribution to the electric and magnetic dipole moments, respectively; $\underline{\mathbf{Q}}_E^e$, $\underline{\mathbf{Q}}_E^m$ are the electric and magnetic quadrupole moment contributions; $\mathbf{P}'_E$ is the third-order contribution to the electric dipole moment. The subscript $E$ for all these quantities indicates the microscopic electrical origin of these multipole moments, all stemming from the *electric* polarization $\mathbf{P}(\mathbf{r})$. We can apply analogous considerations to the microscopic induced magnetization $\mathbf{M}(\mathbf{r})$

$$\overline{\mathbf{M}} = \mathbf{M}_H + \frac{\boldsymbol{\beta}\times\mathbf{P}_H}{\omega\varepsilon_0} - \frac{i\boldsymbol{\beta}}{2}\cdot\underline{\mathbf{Q}}_H^m + \frac{\boldsymbol{\beta}}{i\omega}\times\left[\mathbf{M}'_H\times\boldsymbol{\beta}+\frac{\boldsymbol{\beta}\cdot\underline{\mathbf{Q}}_H^e}{2}\right]+\ldots, \qquad (6)$$

with analogous definitions for the corresponding multipole moments:

$$\mathbf{M}_H = \frac{1}{d^3}\int_V \mathbf{M}(\mathbf{r})d\mathbf{r}$$

$$\mathbf{P}_H = \frac{i\omega\varepsilon_0}{d^3}\int_V \frac{\mathbf{r}\times\mathbf{M}(\mathbf{r})}{2}d\mathbf{r}$$

$$\underline{\mathbf{Q}}_H^m = \frac{1}{d^3}\int_V \left[\mathbf{r}\mathbf{M}(\mathbf{r}) + \mathbf{M}(\mathbf{r})\mathbf{r}\right]d\mathbf{r} \qquad , \qquad (7)$$

$$\mathbf{M}'_H = -\frac{i\omega}{d^3}\int_V \frac{\mathbf{r}\times\mathbf{M}(\mathbf{r})\times\mathbf{r}}{6}d\mathbf{r}$$

$$\underline{\mathbf{Q}}_H^e = \frac{i\omega}{d^3}\int_V \frac{[\mathbf{r}\times\mathbf{M}(\mathbf{r})]\mathbf{r} + \mathbf{r}[\mathbf{r}\times\mathbf{M}(\mathbf{r})]}{3}d\mathbf{r}$$

and the subscript $H$ refers to the microscopic *magnetic* origin of these multipole moments. It should be stressed that for dielectric-only or conducting metamaterials the quantities in (7) are all zero, since $\mathbf{M}(\mathbf{r}) = \mathbf{0}$. This does not mean that magnetic effects are excluded, as the rotation of $\mathbf{P}(\mathbf{r})$ can produce artificial magnetic effects captured by $\mathbf{M}_E$ in (5). Dual considerations apply to magnetic-only inclusions.

Using the previous expansions, in the general magnetodielectric case we may write Eq. (3) as

$$i\boldsymbol{\beta}\times\left[\overline{\mathbf{E}} - \frac{\mathbf{P}_H}{\varepsilon_0} - i\boldsymbol{\beta}\times\mathbf{M}'_H + \frac{i\boldsymbol{\beta}}{2}\cdot\underline{\mathbf{Q}}_H^e\right] = i\omega\left(\mu_0\overline{\mathbf{H}} + \mathbf{M}_H - \frac{i\boldsymbol{\beta}}{2}\cdot\underline{\mathbf{Q}}_H^m\right) - \mathbf{K}_{ext}$$

$$i\boldsymbol{\beta}\times\left[\overline{\mathbf{H}} - \frac{\mathbf{M}_E}{\mu_0} - i\boldsymbol{\beta}\times\mathbf{P}'_E - \frac{i\boldsymbol{\beta}}{2}\cdot\underline{\mathbf{Q}}_E^m\right] = -i\omega\left[\varepsilon_0\overline{\mathbf{E}} + \mathbf{P}_E - \frac{i\boldsymbol{\beta}}{2}\cdot\underline{\mathbf{Q}}_E^e\right] + \mathbf{J}_{ext}$$

$$(8)$$

in which we have neglected the effects of higher order multipole moments (beyond the electric and magnetic quadrupole moments) in Eqs. (4),(6).

*b) Definition of averaged fields*

Eq. (8) ensures that, by correcting the definition of the averaged fields as

$$\mathbf{E}_{av} = \bar{\mathbf{E}} - \frac{\mathbf{P}_H}{\varepsilon_0} - i\mathbf{\beta} \times \mathbf{M}'_H + \frac{i\mathbf{\beta}}{2} \cdot \underline{\mathbf{Q}}^e_H + ...$$

$$\mathbf{H}_{av} = \bar{\mathbf{H}} - \frac{\mathbf{M}_E}{\mu_0} - i\mathbf{\beta} \times \mathbf{P}'_E - \frac{i\mathbf{\beta}}{2} \cdot \underline{\mathbf{Q}}^m_E + ...$$

$$\mathbf{D}_{av} = \varepsilon_0 \bar{\mathbf{E}} + \mathbf{P}_E - \frac{i\mathbf{\beta}}{2} \cdot \underline{\mathbf{Q}}^e_E + ...$$

$$\mathbf{B}_{av} = \mu_0 \bar{\mathbf{H}} + \mathbf{M}_H - \frac{i\mathbf{\beta}}{2} \cdot \underline{\mathbf{Q}}^m_H + ...$$

(9)

the macroscopic (averaged) Maxwell's equations take the expected usual form

$$i\mathbf{\beta} \times \mathbf{E}_{av} = i\omega \mathbf{B}_{av} - \mathbf{K}_{ext}$$
$$i\mathbf{\beta} \times \mathbf{H}_{av} = -i\omega \mathbf{D}_{av} + \mathbf{J}_{ext}$$

(10)

*for any arbitrary pair* $(\omega, \mathbf{\beta})$. Different from the simple spatial averaging in (3), this averaging solves the issues outlined above and ensures the proper representation of artificial electric and magnetic effects, making sure that the constitutive relations tend to local parameters in the long-wavelength limit, even in the presence of artificial magnetic or polarization effects. This averaging procedure, based on a rigorous first-principle approach, constitutes a general framework that properly takes into account weak forms of spatial dispersion associated with artificial magnetism and polarization, at the basis of common metamaterial effects, and it allows their description in a local sense in the long-wavelength limit. Eq. (9) shows that the proper expression for averaged electric and magnetic fields $\mathbf{E}_{av}$ and $\mathbf{H}_{av}$ is obtained after correcting the spatial averages $\bar{\mathbf{E}}$ and $\bar{\mathbf{H}}$ for the possible presence of these artificial effects, associated with the rotation of $\mathbf{P}(\mathbf{r})$ and $\mathbf{M}(\mathbf{r})$. This ensures that these effects are correctly attributed to local constitutive parameters (permeability and permittivity, respectively) in the long-wavelength limit. In the special case of dielectric-only and conducting inclusions $\mathbf{M}(\mathbf{r}) = \mathbf{0}$, $\mathbf{E}_{av} = \bar{\mathbf{E}}$ and $\mathbf{B}_{av} = \mu_0 \bar{\mathbf{H}}$, ensuring that $\mathbf{E}(\mathbf{r})$ and $\mathbf{B}(\mathbf{r})$ are the direct *source* fields, consistent with the homogenization of optical crystals and dielectric metamaterials [33],[23]. Still, instead of defining a generalized permittivity tensor $\underline{\varepsilon}_g$, as in [23], the definition of average magnetic field $\mathbf{H}_{av}$ (9) ensures that artificial magnetic effects are correctly associated to local permeability, consistent with [34]. Conversely, if only magnetization is present at the microscopic level $\mathbf{P}(\mathbf{r}) = \mathbf{0}$, then $\mathbf{H}_{av} = \overline{\mathbf{H}}$ and $\mathbf{D}_{av} = \varepsilon_0 \overline{\mathbf{E}}$ are the source fields, as considered and discussed in [25]. Eq. (9) represents a generalization of these two scenarios to the general case of magnetodielectric inclusions, for which both averaged fields $\mathbf{E}_{av}$, $\mathbf{H}_{av}$ need to be corrected for the possible presence of artificial electric and magnetic effects, respectively. This is the only way to ensure that these weak spatial dispersion effects are properly taken into account within a homogenized description that converges to a local model in the long-wavelength limit.

*c) Relations between averaged fields in the long-wavelength limit*

In the general case, the constitutive relations among the averaged displacement vectors $\mathbf{D}_{av}$, $\mathbf{B}_{av}$ and the averaged field vectors $\mathbf{E}_{av}$, $\mathbf{H}_{av}$ explicitly depend on $\boldsymbol{\beta}$, implying that strong spatial dispersion effects associated with higher-order multipole contributions may be in general present. In the long-wavelength limit of interest here, however, it is expected that the distributions of $\mathbf{P}(\mathbf{r})$, $\mathbf{M}(\mathbf{r})$ may be described exclusively in terms of electric and magnetic dipole moments, which is the case when the unit cell is sufficiently smaller than the wavelength of operation, and the inclusions are not too densely packed. In such circumstances, the explicit effects of spatial dispersion disappear in Eq. (9):

$$\begin{aligned}
\mathbf{E}_{av} &= \bar{\mathbf{E}} - \mathbf{P}_H / \varepsilon_0 \\
\mathbf{H}_{av} &= \bar{\mathbf{H}} - \mathbf{M}_E / \mu_0 \\
\mathbf{D}_{av} &= \varepsilon_0 \bar{\mathbf{E}} + \mathbf{P}_E \\
\mathbf{B}_{av} &= \mu_0 \bar{\mathbf{H}} + \mathbf{M}_H
\end{aligned} \quad (11)$$

and the constitutive model relating averaged displacement and field vectors may be written in the local form

$$\begin{aligned}
\mathbf{D}_{av} &= \varepsilon_0 \mathbf{E}_{av} + \mathbf{P}_E + \mathbf{P}_H = \varepsilon_0 \mathbf{E}_{av} + \mathbf{P}_{av} \\
\mathbf{B}_{av} &= \mu_0 \mathbf{H}_{av} + \mathbf{M}_H + \mathbf{M}_E = \mu_0 \mathbf{H}_{av} + \mathbf{M}_{av}
\end{aligned} \quad (12)$$

where we have combined averaged polarization and magnetization stemming from electric and magnetic microscopic effects into $\mathbf{P}_{av}$ and $\mathbf{M}_{av}$ [39].

Combining Eq. (12) and (2) into (10), we can write a general relation between averaged and external fields and averaged polarization and magnetization vectors,

$$\begin{aligned}
i\boldsymbol{\beta} \times (\mathbf{E}_{av} - \mathbf{E}_{ext}) &= i\omega\mu_0 (\mathbf{H}_{av} - \mathbf{H}_{ext}) + i\omega \mathbf{M}_{av} \\
i\boldsymbol{\beta} \times (\mathbf{H}_{av} - \mathbf{H}_{ext}) &= -i\omega\varepsilon_0 (\mathbf{E}_{av} - \mathbf{E}_{ext}) - i\omega \mathbf{P}_{av}
\end{aligned} \quad (13)$$

These equations may be further manipulated to yield

$$\begin{aligned}
\left[ k_0^2 + \boldsymbol{\beta} \times \boldsymbol{\beta} \times \right] (\mathbf{E}_{av} - \mathbf{E}_{ext}) &= -k_0^2 \frac{\mathbf{P}_{av}}{\varepsilon_0} + k_0 \eta_0 \boldsymbol{\beta} \times \frac{\mathbf{M}_{av}}{\mu_0} \\
\left[ k_0^2 + \boldsymbol{\beta} \times \boldsymbol{\beta} \times \right] (\mathbf{H}_{av} - \mathbf{H}_{ext}) &= -k_0^2 \frac{\mathbf{M}_{av}}{\mu_0} - \frac{k_0}{\eta_0} \boldsymbol{\beta} \times \frac{\mathbf{P}_{av}}{\varepsilon_0}
\end{aligned} \quad (14)$$

where $k_0 = \omega\sqrt{\varepsilon_0 \mu_0}$ and $\eta_0 = \sqrt{\mu_0 / \varepsilon_0}$.

Henceforth, for simplicity of notation we consider only averaged and impressed field distributions that are transverse-electromagnetic (TEM) waves propagating along the direction $\hat{\boldsymbol{\beta}}$ (where the hat indicates a unit vector) [40]. A more general tensorial notation may be used for arbitrary propagation, but is not adopted here in the interest of notational simplicity. In this situation, Eq. (14) may be compactly written as follows,

$$\mathbf{E}_{av} = \mathbf{E}_{ext} + \frac{k_0^2}{\beta^2 - k_0^2}\frac{\mathbf{P}_{av}}{\varepsilon_0} - \frac{\beta k_0}{\beta^2 - k_0^2}\eta_0 \hat{\boldsymbol{\beta}} \times \frac{\mathbf{M}_{av}}{\mu_0}$$
$$\mathbf{H}_{av} = \mathbf{H}_{ext} + \frac{k_0^2}{\beta^2 - k_0^2}\frac{\mathbf{M}_{av}}{\mu_0} + \frac{\beta k_0}{\beta^2 - k_0^2}\frac{\hat{\boldsymbol{\beta}}}{\eta_0} \times \frac{\mathbf{P}_{av}}{\varepsilon_0} ,$$

(15)

where $\beta = |\boldsymbol{\beta}|$. This is a very general result, which relates averaged and impressed fields for *any arbitrary* $(\omega, \boldsymbol{\beta})$ pair and holds for *any* metamaterial array and *any* combination of electric and magnetic excitations. Observe that, similar to the way both electric and magnetic induced currents contribute to electric and magnetic averaged polarization in (12), both averaged polarization and magnetization vectors contribute to averaged electric and magnetic fields. In other words, an inherent form of *magnetoelectric coupling at the unit cell level* stems from weak spatial dispersion effects when $\beta \neq 0$, associated with finite phase velocity across each unit cell. These effects are neglected in quasi-static homogenization methods.

Eq. (15) defines a general relation among averaged and impressed fields, which is independent of the specific nature of the metamaterial inclusions. In the following section, we introduce the inclusion into the picture, and we use this result to define the first-principle effective constitutive model of an arbitrary metamaterial array.

### 3. *Effective* Constitutive Parameters

After having established the proper definition of averaged fields and their general relations, we are now ready to derive a macroscopic homogenized description of the array, once we relate averaged polarization and magnetization vectors to the *local* fields, as a function of the specific inclusion geometry. Since we are assuming that dipolar terms are dominant [39], we may compactly describe the unit cell response in terms of its polarizability coefficients, which relate the induced electric and magnetic dipole moments in the unit cell $\mathbf{p}_{000} = d^3 \mathbf{P}_{av}$ and $\mathbf{m}_{000} = d^3 \mathbf{M}_{av}$ to the *local* fields at its center,

$$\begin{aligned} \mathbf{p}_{000} &= \varepsilon_0 \alpha_e \mathbf{E}_{loc} - \varepsilon_0 \alpha_{em} \eta_0 \hat{\boldsymbol{\beta}} \times \mathbf{H}_{loc} \\ \mathbf{m}_{000} &= \mu_0 \alpha_m \mathbf{H}_{loc} - \mu_0 \alpha_{em} \frac{\hat{\boldsymbol{\beta}} \times \mathbf{E}_{loc}}{\eta_0} \end{aligned}, \quad (16)$$

where $\alpha_e$, $\alpha_m$ and $\alpha_{em}$ are the electric, magnetic and magnetoelectric polarizability coefficients, respectively. All these coefficients have dimensions of an inverse volume, and they are considered scalar here due to the assumptions of TEM propagation and isotropic unit cell. In addition, in writing Eq. (16) we have implicitly assumed that the inclusions are reciprocal [41]. The fields $\mathbf{E}_{loc}$ and $\mathbf{H}_{loc}$ represent the *local* fields at the unit cell center in absence of the inclusion. They are due to the superposition of the impressed fields $\mathbf{E}_{ext}$, $\mathbf{H}_{ext}$ and the induced fields scattered from the rest of the array:

$$\begin{aligned} \mathbf{E}_{loc} &= \mathbf{E}_{array} + \mathbf{E}_{ext} = C \frac{\mathbf{p}_{000}}{\varepsilon_0} - \eta_0 C_{em} \hat{\boldsymbol{\beta}} \times \frac{\mathbf{m}_{000}}{\mu_0} + \mathbf{E}_{ext} \\ \mathbf{H}_{loc} &= \mathbf{H}_{array} + \mathbf{H}_{ext} = C \frac{\mathbf{m}_{000}}{\mu_0} + \frac{C_{em}}{\eta_0} \hat{\boldsymbol{\beta}} \times \frac{\mathbf{p}_{000}}{\varepsilon_0} + \mathbf{H}_{ext} \end{aligned}. \quad (17)$$

The interaction constants $C$ and $C_{em}$ may be evaluated using the dipolar radiation from the generic unit cell at $\mathbf{r}_{lmn} = (ld, md, nd)$ and applying the Floquet condition $\mathbf{p}_{lmn} = \mathbf{p}_{000} e^{i \boldsymbol{\beta} \cdot \mathbf{r}_{lmn}}$, $\mathbf{m}_{lmn} = \mathbf{m}_{000} e^{i \boldsymbol{\beta} \cdot \mathbf{r}_{lmn}}$:

$$C = \sum_{(l,m,n)\neq(0,0,0)} \hat{\mathbf{p}} \cdot \underline{\mathbf{G}}_{ee}(\mathbf{r}_{lmn}) e^{i\boldsymbol{\beta}\cdot\mathbf{r}_{lmn}} \cdot \hat{\mathbf{p}}$$
$$C_{em} = \sum_{(l,m,n)\neq(0,0,0)} \hat{\mathbf{p}} \cdot \underline{\mathbf{G}}_{em}(\mathbf{r}_{lmn}) e^{i\boldsymbol{\beta}\cdot\mathbf{r}_{lmn}} \cdot \hat{\mathbf{m}}$$ (18)

where $\underline{\mathbf{G}}_{ee}(\mathbf{r}_{lmn})$ and $\underline{\mathbf{G}}_{em}(\mathbf{r}_{lmn})$ are the electric and magnetoelectric dyadic Green's functions [43] and $\hat{\mathbf{p}}$, $\hat{\mathbf{m}}$ are unit vectors oriented along $\mathbf{p}_{000}$ and $\mathbf{m}_{000}$, respectively. Fast converging expressions for these summations are available in [15]-[16],[20].

Combining Eqs. (16)-(17), we may now derive a general relation between impressed fields and averaged polarization vectors:

$$\mathbf{E}_{ext} = d^3\left(\frac{1}{\alpha_e + \alpha_{em}^2/\alpha_m} - C\right)\frac{\mathbf{P}_{av}}{\varepsilon_0} + \eta_0 d^3\left(\frac{\alpha_{em}}{\alpha_e\alpha_m + \alpha_{em}^2} + C_{em}\right)\hat{\boldsymbol{\beta}} \times \frac{\mathbf{M}_{av}}{\mu_0}$$
$$\mathbf{H}_{ext} = d^3\left(\frac{1}{\alpha_m + \alpha_{em}^2/\alpha_e} - C\right)\frac{\mathbf{M}_{av}}{\mu_0} + \frac{d^3}{\eta_0}\left(\frac{\alpha_{em}}{\alpha_e\alpha_m + \alpha_{em}^2} - C_{em}\right)\hat{\boldsymbol{\beta}} \times \frac{\mathbf{P}_{av}}{\varepsilon_0}$$ (19)

which, substituted in (15), provides the important relations

$$\mathbf{E}_{av} = \left[\frac{d^3\alpha_m}{\alpha_e\alpha_m + \alpha_{em}^2} - d^3 C_{int}\right]\frac{\mathbf{P}_{av}}{\varepsilon_0} + \left[\frac{d^3\alpha_{em}}{\alpha_e\alpha_m + \alpha_{em}^2} + d^3 C'_{em}\right]\eta_0\hat{\boldsymbol{\beta}} \times \frac{\mathbf{M}_{av}}{\mu_0}$$
$$\mathbf{H}_{av} = \left[\frac{d^3\alpha_e}{\alpha_e\alpha_m + \alpha_{em}^2} - d^3 C_{int}\right]\frac{\mathbf{M}_{av}}{\mu_0} + \left[\frac{d^3\alpha_{em}}{\alpha_e\alpha_m + \alpha_{em}^2} - d^3 C'_{em}\right]\frac{\hat{\boldsymbol{\beta}}}{\eta_0} \times \frac{\mathbf{P}_{av}}{\varepsilon_0}$$ (20)

Here we have used the reduced interaction constants

$$C_{int} = C - \left[\frac{1}{d^3}\frac{k_0^2}{\beta^2 - k_0^2}\right]$$
$$C'_{em} = C_{em} - \left[\frac{1}{d^3}\frac{\beta k_0}{\beta^2 - k_0^2}\right]$$ (21)

which respectively coincide with $\hat{\mathbf{p}} \cdot \underline{\mathbf{C}}_{int} \cdot \hat{\mathbf{p}}$ and $\hat{\mathbf{p}} \cdot \underline{\mathbf{C}}_{e,m} \cdot \hat{\mathbf{m}}$ derived in [23] using an alternative spectral-domain representation.

Eq. (20) represents another important result, since it shows, directly from first-principle considerations, that it is possible to establish a general relation between averaged electric and magnetic fields and averaged polarization vectors [as defined in (11)], which depends on the array period and the polarizability coefficients for any given pair $(\omega, \boldsymbol{\beta})$, but that is *completely independent* of the relative amplitude of the impressed sources $\mathbf{J}_{ext}$, $\mathbf{K}_{ext}$. This is a relevant property of this homogenization theory, since a proper homogenized description of the metamaterial should not depend on the type and form of external excitation, as commonly happens in more approximate homogenization models.

The relations (20) also show that there is an inherent form of *magnetoelectric* coupling (usually negligible in natural materials) relating $\mathbf{E}_{av}$ to the rotation of $\mathbf{M}_{av}$, and $\mathbf{H}_{av}$ to the rotation of $\mathbf{P}_{av}$. As expected, part of this coupling is associated with the presence of $\alpha_{em}$, which represents the possible bianisotropy within the unit cell, stemming from asymmetric or noncentered inclusions [44]. However, Eq. (20) predicts that, even when inclusions are perfectly centersymmetric and with no inherent bianisotropy, a form of magnetoelectric coupling is still expected, associated with the presence of $C'_{em}$. This additional coupling term is due to lattice effects and the nonzero value of $\beta$. We will discuss the implications of this coupling in more detail in the following.

Using (12), we can finally write for the constitutive relations of the metamaterial array:

$$\begin{aligned}\mathbf{D}_{av} &= \varepsilon_0 \mathbf{E}_{av} + \mathbf{P}_{av} = \varepsilon_{eff} \mathbf{E}_{av} - \left(\chi^e_{eff} + \chi^o_{eff}\right)\hat{\boldsymbol{\beta}} \times \mathbf{H}_{av} \\ \mathbf{B}_{av} &= \mu_0 \mathbf{H}_{av} + \mathbf{M}_{av} = \mu_{eff} \mathbf{H}_{av} - \left(\chi^e_{eff} - \chi^o_{eff}\right)\hat{\boldsymbol{\beta}} \times \mathbf{E}_{av}\end{aligned} \quad (22)$$

with

$$\varepsilon_{eff} = \varepsilon_0 \left[ 1 + \frac{d^{-3}\left[\alpha_{m(eff)}^{-1} - C_{int}\right]}{\left(\alpha_{e(eff)}^{-1} - C_{int}\right)\left(\alpha_{m(eff)}^{-1} - C_{int}\right) - C_{em}'^2 + \alpha_{em(eff)}^2} \right]$$

$$\mu_{eff} = \mu_0 \left[ 1 + \frac{d^{-3}\left(\alpha_{e(eff)}^{-1} - C_{int}\right)}{\left(\alpha_{e(eff)}^{-1} - C_{int}\right)\left(\alpha_{m(eff)}^{-1} - C_{int}\right) - C_{em}'^2 + \alpha_{em(eff)}^2} \right], \quad (23)$$

$$\chi_{eff}^e = \frac{1}{c_0} \frac{d^{-3}\alpha_{em(eff)}}{\left(\alpha_{m(eff)}^{-1} - C_{int}\right)\left(\alpha_{m(eff)}^{-1} - C_{int}\right) - C_{em}'^2 + \alpha_{em(eff)}^2}$$

$$\chi_{eff}^o = \frac{1}{c_0} \frac{d^{-3}C_{em}'}{\left(\alpha_{e(eff)}^{-1} - C_{int}\right)\left(\alpha_{m(eff)}^{-1} - C_{int}\right) - C_{em}'^2 + \alpha_{em(eff)}^2}$$

where $c_0 = 1/\sqrt{\varepsilon_0\mu_0}$, $\alpha_{e(eff)}^{-1} = \alpha_m/\Delta$, $\alpha_{m(eff)}^{-1} = \alpha_e/\Delta$, $\alpha_{em(eff)}^{-1} = \alpha_{em}/\Delta$ and $\Delta = \left(\alpha_e\alpha_m + \alpha_{em}^2\right)$ [in the absence of magnetoelectric coupling at the unit cell level $\alpha_{em} = 0$ and $\alpha_{e(eff)} = \alpha_e$, $\alpha_{m(eff)} = \alpha_m$].

*a) General properties of the effective constitutive parameters*

The expressions (23) represent general closed-form effective constitutive parameters, rigorously obtained from first-principle considerations. They are valid for any pair $(\omega, \boldsymbol{\beta})$ and any form of external excitation $\mathbf{J}_{ext}$, $\mathbf{K}_{ext}$, ensuring that this homogenized description does not depend on the specific impressed field distribution in each unit cell, but instead represents the inherent response of the metamaterial as a bulk to an arbitrary electric and/or magnetic excitation. It is important to stress that the ratio of averaged fields $E_{av}/H_{av}$, i.e., the local wave impedance, inherently depends on the specific choice of impressed sources $\mathbf{J}_{ext}$, $\mathbf{K}_{ext}$ as in Eq. (13) and as expected in presence of arbitrarily impressed sources. However, the constitutive parameters defined in Eq. (23) do not depend on this ratio, thus compactly describing the macroscopic polarization and magnetization properties of the array *for arbitrary excitation*. This fundamental property does not apply to less rigorous homogenization models that focus only on eigenmodal propagation, as we discuss more extensively in Section 4.

Effective permittivity and permeability are found in closed form as the first two expressions (23). These generalize the Clausius-Mosotti homogenization formulas [6]-[7],[23] by rigorously taking into account the coupling among the inclusions and their polarization properties. More importantly, this theory demonstrates the inherent presence of magnetoelectric coupling via the coefficients $\chi_{eff}^{e}$ and $\chi_{eff}^{o}$ in Eq. (22)-(23). The first portion of the bianisotropy coefficient $\chi_{eff}^{e}$, even with respect to $\beta$, is associated with magnetoelectric effects at the inclusion level, and satisfies the usual reciprocity constraints for bulk materials. An additional contribution to bianisotropy is $\chi_{eff}^{o}$, odd with respect to $\beta$, which is associated to inherent magnetoelectric coupling effects arising at the lattice level. These latter effects cannot be neglected in general, even in the case of center-symmetric inclusions for which $\alpha_{em} = 0$, as long as $C'_{em} \neq 0$. The presence of this odd bianisotropy coefficient has been pointed out theoretically and numerically in [21]-[22], and the present theory explains its physical nature and relevance from first-principle considerations: $\chi_{eff}^{o}$ is a weak spatial dispersion effect associated with the finite phase velocity along the array, not negligible even in the long-wavelength limit as we show in the following numerical examples (Section 7). Its nature is associated with the inherent asymmetry introduced by phase propagation across a unit cell of finite size, and this explains why, at first sight, its occurrence in (22) does not satisfy the reciprocity relation for local bianisotropic materials. Its odd response with respect to $\beta$ ensures that, by reversing the direction of propagation for given frequency, its contribution also changes sign, ensuring that the constitutive relations (22) are actually describing a reciprocal medium. This shows the drastic difference between the bianisotropy stemming from lattice effects $\chi_{eff}^{o}$ and the one associated with magnetoelectric coupling at the inclusion level $\chi_{eff}^{e}$. Its relevance in the homogenization of metamaterials and in restoring the physical meaning of their constitutive parameters is discussed in further detail in [32].

Due to the inherent properties of the summations in (18), which for real $\beta$ satisfy [15]-[16],[20],[23]

$$\text{Im}[C] = k_0^3 / (6\pi)$$
$$\text{Im}[C_{em}] = 0 \qquad (24)$$

and the lossless conditions on the polarizability coefficients [45]

$$\text{Im}\left[\alpha_e^{-1}\right] = \text{Im}\left[\alpha_m^{-1}\right] = k_0^3 / (6\pi)$$
$$\text{Im}[\alpha_{em}] = 0 \qquad (25)$$

it is recognized that all the constitutive parameters in (23) are real for lossless particles and real $\beta$, as required in lossless bianisotropic materials, ensuring power conservation.

Before concluding this section, it is worth stressing that the closed-form expressions (23) apply to any plane-wave like field distribution in the homogenized material, any form of excitation, and any pair $(\omega, \boldsymbol{\beta})$, representing an accurate and self-consistent homogenization model for the array. The constitutive parameters may still be, in the general case, weakly dependent on $\beta$, as a symptom of spatial dispersion. However, as discussed in Section 5, the homogenized parameters tend to a local, nondispersive model in the long-wavelength limit (small $\beta$). The present theory proves that it is possible to rigorously derive from first-principles a self-consistent constitutive model for metamaterials. In addition to artificial magnetism and polarization effects stemming from weak spatial dispersion, this theory shows that the rigorous consideration of the array coupling requires an additional magnetoelectric constitutive parameter, even in the case of centersymmetric inclusions. In the following section, we consider the special case of eigenmodal solution (without impressed sources), and relate our general theory to other common eigenmodal approaches to homogenization.

## 4. Eigenmodal Propagation and *Equivalent* Constitutive Parameters

In the eigenmodal case, i.e., in the absence of external sources, Eq. (19) ensures that non-trivial solutions are available only for specific instances of $\boldsymbol{\beta}(\omega)$, satisfying the array dispersion relation

$$\left(\alpha_{e(eff)}^{-1} - C\right)\left(\alpha_{m(eff)}^{-1} - C\right) = C_{em}^2 - \alpha_{em(eff)}^{-2}. \tag{26}$$

The corresponding eigenvectors, obtained solving Eq. (19), satisfy

$$\frac{\mathbf{p}_{000} \cdot \hat{\mathbf{p}}}{\mathbf{m}_{000} \cdot \hat{\mathbf{m}}} = \frac{P_{av}}{M_{av}} = \frac{1}{\eta_0} \frac{C_{em} + \alpha_{em(eff)}}{\alpha_{e(eff)}^{-1} - C} = \frac{1}{\eta_0} \frac{\alpha_{m(eff)}^{-1} - C}{C_{em} - \alpha_{em(eff)}}, \tag{27}$$

which provides a specific constraint on the ratio $P_{av}/M_{av}$. Rearranging Eq. (22) and (13), in this regime we may also write

$$\begin{aligned}
i\boldsymbol{\beta} \times \mathbf{E}_{av} &= i\omega \frac{\mu_{eff}}{1 - \dfrac{c_0\left(\chi_{eff}^o - \chi_{eff}^e\right)}{\beta/k_0}} \mathbf{H}_{av} = i\omega \mu_{eq} \mathbf{H}_{av} \\
i\boldsymbol{\beta} \times \mathbf{H}_{av} &= -i\omega \frac{\varepsilon_{eff}}{1 - \dfrac{c_0\left(\chi_{eff}^o + \chi_{eff}^e\right)}{\beta/k_0}} \mathbf{E}_{av} = -i\omega \varepsilon_{eq} \mathbf{E}_{av}
\end{aligned}, \tag{28}$$

where $c_0$ is the velocity of light in free-space. Eq. (28) shows that the eigenmodal propagation may be described in terms of *equivalent* permittivity and permeability parameters $\varepsilon_{eq}$, $\mu_{eq}$, which embed the magnetoelectric coupling effects as a form of weak spatial dispersion. Their validity is strictly limited to eigenmodal propagation, since the ratio $P_{av}/M_{av}$ is in general a function of the impressed sources. The description of the array in terms of equivalent parameters is particularly attractive in absence of bianisotropic effects at the inclusion level, when $\alpha_{em} = \chi^e_{eff} = 0$. In this case, the residual magnetoelectric coupling associated with lattice effects may be embedded into *equivalent* permittivity and permeability parameters related to the *effective* parameters through the normalization factor $1 - \dfrac{c_0 \chi^o_{eff}}{\beta / k_0}$.

Classic homogenization models that aim at describing metamaterial arrays in terms of permittivity and permeability (see, e.g., [13]-[20]), on the model of natural materials, extract and define these *equivalent* quantities, and thus implicitly introduce a form of weak spatial dispersion when $\chi^o_{eff}$ is not negligible. It is evident that this may easily translate into inconsistencies and lack of physical meaning in the extracted or retrieved parameters, as discussed in more detail in [32], and verified in several recent examples in the literature [27]-[28]. It is worth stressing that the *equivalent* parameters are an implicit function of the specific ratio $P_{av}/M_{av}$ in (27), i.e., they are bound to change when impressed sources are introduced that can arbitrarily modify the local ratio $P_{av}/M_{av}$, in sharp contrast with the general independence of the effective constitutive parameters (23) on the local value of $P_{av}/M_{av}$.

*a) Secondary parameters and relations between equivalent and effective descriptions*

It follows straightforwardly from (28) that the dispersion relation $\boldsymbol{\beta}(\omega)$ may be rewritten as

$$\beta^2 = \omega^2 \mu_{eq} \varepsilon_{eq}, \tag{29}$$

which, after using Eqs. (23) and (21), may be shown to coincide with Eq. (26).

In addition, we can define the effective characteristic impedance of the array for eigenmodal propagation:

$$\eta_{eff} = \frac{E_{av}}{H_{av}} = \sqrt{\frac{\mu_{eq}}{\varepsilon_{eq}}} = \sqrt{\frac{\mu_{eff}}{\varepsilon_{eff}}} \sqrt{\frac{\beta/k_0 - c_0\left(\chi_{eff}^o + \chi_{eff}^e\right)}{\beta/k_0 - c_0\left(\chi_{eff}^o - \chi_{eff}^e\right)}}. \qquad (30)$$

In the absence of bianisotropic effects in the inclusions $\alpha_{em} = \chi_{eff}^e = 0$ and Eq. (30) becomes

$$\eta_{eff} = \frac{E_{av}}{H_{av}} = \sqrt{\frac{\mu_{eq}}{\varepsilon_{eq}}} = \sqrt{\frac{\mu_{eff}}{\varepsilon_{eff}}}. \qquad (31)$$

Therefore, in absence of bianisotropy the eigenmodal characteristic impedance is not directly affected by magnetoelectric coupling at the lattice level, and the same characteristic impedance is obtained using either the ratio of *effective* or *equivalent* parameters. In addition, using Eq. (22) we may write in the general case

$$\frac{P_{av}}{M_{av}} = \frac{\left(\varepsilon_{eff} - \varepsilon_0\right)\eta_{eff} + \left(\chi_{eff}^e + \chi_{eff}^o\right)}{\left(\mu_{eff} - \mu_0\right) - \eta_{eff}\left(\chi_{eff}^e - \chi_{eff}^o\right)}, \qquad (32)$$

which, for $\alpha_{em} = \chi_{eff}^e = 0$, becomes

$$\frac{P_{av}}{M_{av}} = \eta_{eff} \frac{\varepsilon_{eq} - \varepsilon_0}{\mu_{eq} - \mu_0}. \qquad (33)$$

Eqs. (29) and (33) coincide with classic retrieval procedures used to determine the effective permittivity and permeability of a metamaterial sample from its secondary parameters, i.e., its eigenmodal wave number $\beta$ and its characteristic impedance $\eta_{eff}$ [20],[29]. This means that the *equivalent* representation (28), introduced here from first-principles, exactly coincides with classic homogenization models based on retrieved parameters.

This is a salient finding, since it allows us to relate the present theory to classic homogenization schemes and shows that simple constitutive models assigned *a priori* may effectively hide weak spatial dispersion effects. In source-free problems, for which the excitation is placed outside the metamaterial sample, as in classic retrieval schemes, it is tempting to put aside the magnetoelectric coupling coefficient $\chi_{eff}^{o}$, and use the *equivalent* parameters to model the array scattering. This is indeed possible, and from the scattering point of view *effective* and *equivalent* descriptions are equivalent in this source-free scenario, since the corresponding secondary parameters coincide. However, our theory shows that the *equivalent* representation, so common in standard metamaterial homogenization schemes, has a very limited physical meaning and it should not be used to separately describe the electric and magnetic response of a metamaterial, since it hides an inherent form of spatial dispersion and magnetoelectric coupling when $\chi_{eff}^{o}$ is not negligible. It is not surprising that the frequency dispersion of the *equivalent* parameters may contain nonphysical artifacts and may not satisfy passivity, reciprocity or other causality constraints typical of local parameters [32].

As a final remark with respect to standard homogenization schemes, the relation between the equivalent parameters and classic retrieval techniques shows that, even at frequencies where spatial dispersion and $\chi_{eff}^{o}$ are negligible and we can safely write

$$\begin{aligned}\mathbf{D}_{av} &= \varepsilon_{eq}\mathbf{E}_{av} \\ \mathbf{B}_{av} &= \mu_{eq}\mathbf{H}_{av}\end{aligned} \quad (34)$$

as in a natural material, the averaged fields $\mathbf{E}_{av}$ and $\mathbf{H}_{av}$ are defined through Eq. (11) and not as the simple spatial averages $\overline{\mathbf{E}}$, $\overline{\mathbf{H}}$ of the microscopic fields. This means that standard retrieval techniques based on the local model (34) implicitly assume:

$$\varepsilon_{eq} = \frac{\varepsilon_0 \mathbf{E}_{av} + \mathbf{P}_{av}}{\mathbf{E}_{av}} = \frac{\varepsilon_0 \overline{\mathbf{E}} + \mathbf{P}_E}{\overline{\mathbf{E}} - \mathbf{P}_H / \varepsilon_0}$$
$$\mu_{eq} = \frac{\mu_0 \mathbf{H}_{av} + \mathbf{M}_{av}}{\mathbf{H}_{av}} = \frac{\mu_0 \overline{\mathbf{H}} + \mathbf{M}_H}{\overline{\mathbf{H}} - \mathbf{M}_E / \mu_0}$$
(35)

The nature of the averaged polarization currents within each unit cell, whether stemming from microscopic electric or magnetic effects, inherently determines the definition of the spatial averages used to calculate the constitutive parameters, and weak spatial dispersion effects associated with artificial magnetism or polarization have a different role (contributing to $\mathbf{E}_{av}$ and $\mathbf{H}_{av}$) than the direct polarization and magnetization vectors (contributing to $\mathbf{D}_{av}$ and $\mathbf{B}_{av}$). Our theory effectively shows that any time we describe metamaterials in terms of permittivity and permeability we implicitly define the averaged fields as in (35) and not by simply taking the spatial averages of the microscopic fields. In particular, $\mathbf{D}_{av}$ and $\mathbf{B}_{av}$ take into account only the direct macroscopic effects $\mathbf{P}_E$ and $\mathbf{M}_H$ of the microscopic polarization and magnetization, respectively. Conversely, the average fields $\mathbf{E}_{av}$ and $\mathbf{H}_{av}$ are implicitly obtained after subtracting the artificial electric and magnetic effects $\mathbf{P}_H$ and $\mathbf{M}_E$ from the spatial averages of $\mathbf{E}(\mathbf{r})$ and $\mathbf{H}(\mathbf{r})$.

## 5. Long-Wavelength Limit and Convergence to a Local Model

The effective parameters $\varepsilon_{eff}$, $\mu_{eff}$ and $\chi_{eff}^e$ in (23) are even functions of $\beta$, as expected from reciprocity considerations. This implies that for $\beta d \ll 1$ they tend to local parameters, one of the relevant advantages of this model, compared to other Floquet approaches to homogenization [23],[33]. In contrast, $\chi_{eff}^o$ is an odd function of $\beta$, which varies linearly with $\beta d$ in the same long-wavelength limit. These considerations imply that the constitutive model (22) converges to the local relations

$$\begin{aligned}\mathbf{D}_{av} &= \varepsilon_{eff}\mathbf{E}_{av} - \chi^e_{eff}\hat{\boldsymbol{\beta}}\times\mathbf{H}_{av} - \kappa_{eff}\boldsymbol{\beta}\times\mathbf{H}_{av}\\ \mathbf{B}_{av} &= \mu_{eff}\mathbf{H}_{av} - \chi^e_{eff}\hat{\boldsymbol{\beta}}\times\mathbf{E}_{av} + \kappa_{eff}\boldsymbol{\beta}\times\mathbf{E}_{av}\end{aligned} \qquad (36)$$

where $\kappa_{eff} = \chi^o_{eff}/\beta$ [32] is also an even function of $\beta$, and all the effective parameters in (36) may be assumed local in the long-wavelength limit. Eq. (36) stresses the relevance of the magnetoelectric coefficient $\kappa_{eff}$ even in this limit, which is one of the relevant results of the present theory, discussed in more details in [32].

For sufficiently long wavelength and away from the inclusion resonances, under the conditions $k_0 d \ll 1$, $\beta d \ll 1$, this lattice effect becomes negligible:

$$C'_{em} \simeq 0, \quad \kappa_{eff} \simeq 0. \qquad (37)$$

Under this simple condition, and in the absence of bianisotropic effects at the inclusion level $\alpha_{em} = 0$, the constitutive parameters (23) become

$$\begin{aligned}\varepsilon_{eff} &= \varepsilon_{eq} = \varepsilon_0\left(1 + \frac{d^{-3}}{\alpha_e^{-1} - C_{int}}\right)\\ \mu_{eff} &= \mu_{eq} = \mu_0\left(1 + \frac{d^{-3}}{\alpha_m^{-1} - C_{int}}\right)\end{aligned} \qquad (38)$$

which coincide with generalized Clausius-Mossotti relations in [23]. Under condition (37) and $\alpha_{em} = 0$, we find for the eigenmodal solution:

$$\begin{aligned}\beta &= \pm\omega\sqrt{\mu_{eff}\varepsilon_{eff}}\\ \frac{\mathbf{p}_{000}\cdot\hat{\mathbf{p}}}{\mathbf{m}_{000}\cdot\hat{\mathbf{m}}} &= \frac{P_{av}}{M_{av}} = \eta_{eff}\frac{\varepsilon_{eff}-\varepsilon_0}{\mu_{eff}-\mu_0},\end{aligned} \qquad (39)$$

which coincides with Eqs. (29) and (33) for equivalent parameters. This proves that $\varepsilon_{eq}$ and $\mu_{eq}$, as well as the retrieved parameters as defined in [20], are identical with $\varepsilon_{eff}$ and $\mu_{eff}$ when (a) there are no impressed sources and (b) magnetoelectric effects at the lattice level are negligible. If, in the very long-wavelength limit, $C_{int}$ may also be assumed independent of $\beta$, then, by Taylor expanding its expression in terms of $k_0$ we get the known approximation [6]

$$C_{int}(\omega, \beta \to 0) = \frac{1}{3d^3} + j\frac{k_0^3}{6\pi}, \quad (40)$$

which proves that Eqs. (38) and this homogenization method both converge to local classic Clausius-Mossotti formulas for periodic arrays [5]-[7] when $\omega, \beta \to 0$. In Section 7, we show that the assumptions (37) and (40) do not necessarily hold in metamaterials, even for $k_0 d < 1$, and that the homogenization approach introduced here may provide results significantly different from quasi-static approaches.

## 6. Spatial Dispersion and Extreme Metamaterial Parameters

It is near the inclusion resonances that metamaterials find the most practical interest, since it is in this frequency range that the constitutive parameters assume extreme (very large, very low, or negative) values. The homogenization model described here is very general, and in principle applicable to any value of $(\omega, \beta)$. However, the same definition of *homogenization* implies the inherent neglect of the array granularity. This is particularly relevant near these resonances, since, despite a small $k_0 d$, the effective eigenmodal wavelength may become comparable with the period as $\beta d$ increases. Although these resonant regions are quite limited in bandwidth for passive inclusions in the long-wavelength regime, it is here that the effects of $\chi_{eff}^{o}$ and spatial dispersion are most relevant.

   *a) Near-zero effective material parameters*

Limiting ourselves to the lossless scenario for clarity, consider first the low-index regime, for which $\beta d \simeq 0$ for finite $k_0 d$, of interest in a variety of applications [46]-[51]. This regime includes ε-near-zero, μ-near-zero and low-index metamaterials. In this frequency range, the eigenwave number $\beta$ passes from being imaginary to real-valued, since one of the two equivalent parameters crosses the real axis [see eq. (29)]. As expected, the effective parameters $\varepsilon_{eff}$, $\mu_{eff}$ and $\chi_{eff}^{e}$ in (23) are real-valued also when $\beta$ is purely imaginary, since they are even functions of $\beta$. On the other hand, $\chi_{eff}^{o}$ is purely imaginary for imaginary $\beta$ and crosses zero for $\beta = 0$, due to its odd nature. This ensures that, when $\chi_{eff}^{e} = 0$ and the *equivalent* representation is appealing, also $\varepsilon_{eq}$ and $\mu_{eq}$ are real-valued (and one of them negative) in Eq. (28), despite $\beta$ being imaginary. As shown in some of the following numerical examples, this zero-index region provides significant deviations between the equivalent parameters $(\varepsilon_{eq}, \mu_{eq})$ and the effective parameters $(\varepsilon_{eff}, \mu_{eff})$, as a symptom of inherent spatial dispersion, consistent with the results in [26]. It should be stressed that in this region $\beta$, $C'_{em}$, $\chi_{eff}$ are all very close to zero, implying very long effective wavelengths and weak magnetoelectric coupling; however, the ratio $\chi_{eff}^{o} / \beta = \kappa_{eff}$ is not necessarily small in the denominator of (28) and in (36), providing relevant nonlocal effects in the equivalent parameters [32].

   *b) Effective parameters near the bandgap regions*

Another region of interest for metamaterial applications is the one near the edge of the lattice bandgaps, for which $\beta d \simeq \pi$. Around this region, large positive or negative values of permittivity and permeability are obtained, of interest in a variety of applications [18], [52]-[53]. It is evident that in this scenario the inclusion interaction may become very complex, and an average over the unit cell may not provide much insight into the physical behavior of an eigenmode that flips its phase within a single unit cell. In particular, inside the bandgap the same definition of homogenized parameters is not meaningful, as they become complex even for lossless inclusions, since $\beta$ is in general complex. It is meaningful, however, to study the transition between the homogenization and the bandgap regimes, where extreme metamaterial parameters are found. It is in this transition region that our homogenization technique becomes particularly important, since here weak spatial dispersion effects as in (36) become relevant, even in the long-wavelength limit $k_0 d < 1$. Exactly at the bandgap edge the periodic properties of $C_{em}$ require that

$$C_{em}(\beta = \pi/d) = 0 \quad \forall \omega. \tag{41}$$

For centersymmetric inclusions ($\alpha_{em} = 0$), this implies that the general dispersion relation (26) simplifies into

$$\left[\alpha_e^{-1} - C(\omega, \pi/d)\right]\left[\alpha_m^{-1} - C(\omega, \pi/d)\right] = 0. \tag{42}$$

In the long-wavelength limit for which $C$ is small, Eq. (42) ensures that a bandgap may be reached exclusively near an electric or a magnetic resonance, for which one of the two $\alpha^{-1} = C \simeq 0$ [54]. It follows directly from (23),(28) that at such resonance one of the equivalent parameters

$$\mu_{eq} = \mu_0 \text{ (for } \alpha_e^{-1} = C\text{) or } \varepsilon_{eq} = \varepsilon_0 \text{ (for } \alpha_m^{-1} = C\text{).} \tag{43}$$

Correspondingly, using (29) the other equivalent parameter has to become

$$\varepsilon_{eq} = \varepsilon_0 \frac{\pi^2}{(k_0 d)^2} \text{ (for } \alpha_e^{-1} = C\text{) or } \mu_{eq} = \mu_0 \frac{\pi^2}{(k_0 d)^2} \text{ (for } \alpha_m^{-1} = C\text{)}. \tag{44}$$

For instance, if we consider the bandgap associated with a magnetic resonance $\alpha_m^{-1} = C$, as is the case for the first resonance of a dielectric inclusion (see example 2 in the following section), the eigenwave number $\beta$ and the corresponding effective permeability $\mu_{eff}$ rapidly increase approaching the bandgap from below. At the resonance $\alpha_m^{-1} = C$, $\varepsilon_{eq} = \varepsilon_0$, $\mu_{eq} = \mu_0 \frac{\pi^2}{(k_0 d)^2}$, and therefore, using (28)

$$\begin{aligned} \frac{\varepsilon_{eff}}{\varepsilon_0} &= 1 - \frac{c_0 \chi_{eff}^o}{\pi/(k_0 d)} \\ \frac{\mu_{eff}}{\mu_0} &= \frac{\pi}{k_0 d}\left(\frac{\pi}{k_0 d} - c_0 \chi_{eff}^o\right) \end{aligned}. \tag{45}$$

Eq. (30) is indeed satisfied by Eq. (45), and it implies

$$\frac{\eta_{eff}}{\eta_0} = \frac{\beta}{k_0} = \frac{\pi}{(k_0 d)}, \tag{46}$$

which suggests that, independent of the inclusion geometry, at a magnetic bandgap edge the normalized characteristic impedance coincides with the normalized wave number (index of refraction). An analogous derivation for electric resonances $\alpha_e^{-1} = C$ provides the inverse of Eq. (46). It is evident that in this regime $\chi_{eff}^o$ may not be neglected and its effect is indeed comparable, if not more important, than the effects captured by $\varepsilon_{eff}$ and $\mu_{eff}$. In this frequency range the *equivalent* parameters (28) lose their physical meaning and strongly diverge from the *effective* parameters (23), as discussed in further detail in [32].

It is evident from this discussion that regions with extreme (very large, very low or negative) metamaterial parameters are those for which the present homogenization technique is most useful, as it diverges from classic homogenization schemes applicable for natural materials and mixtures. It is interesting that a simple local model as in Eq. (36) can capture these effects and fully restore the physical meaning of effective homogenized parameters. We provide numerical examples in Section 7 to illustrate how this rigorous model may correctly capture the exotic features of metamaterials and highlight the weak spatial dispersion effects that are usually at the root of inconsistencies in less rigorous homogenization models.

## 7. Numerical Examples and Further Discussion

In this section, we discuss the homogenization of three specific metamaterial geometries. Although our general formulation is applicable to lossy, bianisotropic, magnetodielectric inclusions, arbitrary source distribution and any choice of $(\omega, \boldsymbol{\beta})$, here we focus on metamaterials composed of lossless dielectric or conducting spheres and on eigenmodal propagation. This choice has the advantage of providing a clearer picture of the difference between this homogenization approach and other available techniques, tailored for eigenmodal excitation. In addition, the choice of center-symmetric inclusions ensures that bianisotropic effects can only stem from the effects captured by $\chi_{eff}^{o}$. We limit our analysis to a dipolar model and long-wavelength regime $(k_0 d) < 1$, usually considered safe for quasi-static homogenization models of metamaterials [9]-[11]. For this reason, we concentrate here on dielectric or conducting inclusions, since their magnetic effects are properly captured by the magnetic polarizability, consistent with the note in [39]. In future works we will apply our general multipolar approach introduced in Section 2 to arbitrary metamaterial inclusions and extend our numerical analysis to the presence of embedded sources and magnetodielectric inclusions [55]. Since we deal with spherical particles, we can use analytical closed-form expressions for $\alpha_e$, $\alpha_m$ [56], well aware of the small causality violations introduced by this assumption, as discussed in [42]. The parameter $\gamma = a/d$ is introduced to define the ratio of sphere radius over lattice period, as a measure of the array density.

Figure 1a shows in logarithmic scale the amplitude of the normalized polarizability coefficients (thick lines for the normalized electric polarizability, thin for the magnetic one) for three different geometries of interest: (1, solid lines) dielectric spheres with relative permittivity $\varepsilon_r = 20$ and permeability $\mu_r = 1$; (2, dashed) dielectric spheres with $\varepsilon_r = 120$ and $\mu_r = 1$; (3, dotted) perfectly conducting spheres. For convenience, Fig. 1b shows the ratio $|\alpha_e|/|\alpha_m|$, in order to highlight the ratio between electric and magnetic response at the inclusion level. Both plots show their variation as a function of $(k_0 d)$, for the density factor $\gamma = 0.45$. The chosen geometries represent specific situations of interest in common metamaterial arrays: in case 1 (solid lines), a regular array of dielectric spheres is considered, far from their individual resonances, but still with a good contrast compared to the background: a dominant electric polarization is expected all over the spectrum of interest; in case 2 (dashed), the permittivity is increased to support a magnetic and an electric resonance within the frequency band of interest, in analogy with established designs to realize negative metamaterial parameters [18]: in this case, more interesting features are expected in the metamaterial response near the inclusion resonances. As expected, the electric response is dominant for lower frequencies, but the first resonance is magnetic. Finally, in case 3 (dotted), conducting particles are considered, for which electric and magnetic responses are comparable, and for lower frequencies the electric polarizability is exactly twice the magnetic one. It is noticed that in all these examples, lossless conditions (25) strictly apply. The application of this theory to magnetodielectric spheres that support negative index of refraction has been considered in [32].

Figure 2a shows the dispersion of normalized eigenwave number (effective index of refraction) for the array 1 with $\gamma = 0.45$. The figure compares the exact eigenmodal solution $\beta / k_0$ (solid line), as obtained from Eq. (26), with various approximate solutions obtained neglecting spatial dispersion and magnetoelectric effects, as follows: the dashed line refers to the dispersion of $\beta_{em}$, obtained neglecting the magnetoelectric coupling term $C'_{em}$, as in Eq. (37); the dotted line shows $\beta_{CM}$, which in addition neglects the dispersion effects in $C_{int}$, implying $C'_{em} = 0$ and $C_{int}$ as given by Eq. (40), coinciding with the quasi-static Clausius-Mossotti homogenization model; the dash-dotted line refers to $\beta_e$, obtained neglecting the magnetic polarizability effects associated with the magnetism of the inclusions (which is small in this geometry), but still using the exact $C_{int}$ expression; finally the dash-dot-dot line refers to $\beta_{e-CM}$, which neglects the magnetic effects and uses Eq. (40) for $C_{int}$. We consider all these approximate expressions to show how the different spatial and frequency dispersion terms, usually neglected in quasi-static homogenization models, affect the metamaterial homogenization accuracy, within the same dipolar approximation. As expected, all these expressions converge to the same quasi-static limit when $(\omega, \beta) \to 0$, but the approximate expressions start deviating from the exact expression of $\beta$ for relatively low values of $k_0 d$. In particular, by neglecting the magnetic polarizability of the particles, which in this example is orders of magnitude smaller than the electric one (see Fig. 1b), the dispersion of $\beta_e / k_0$ surprisingly diverges quite drastically from the exact model, implying that the small magnetism of these dielectric particles cannot be neglected, as one may be tempted after inspecting Fig. 1b. The effects of nonlocality and spatial dispersion in $C_{int}$ start playing a role much earlier in frequency than one would generally expect for such simple topology, comparing $\beta_{CM}$ with $\beta$. In comparison, magnetoelectric coupling effects have a much weaker role, and start being relevant only around $k_0 d \simeq 1$. Figure 2b, in comparison, shows the same curves for the case of a less dense array, with $\gamma = 0.3$. As visible, the trend is quite similar, although effects of spatial dispersion are proportionally less relevant here, as the interaction among inclusions is weaker. In particular, magnetoelectric coupling effects associated with $C'_{em}$ are negligible all over the considered frequency range, as $\beta_{em}$ practically coincides with $\beta$ in this less dense configuration. Nonlocal effects in $C_{int}$ and the influence of the small magnetic properties of the inclusions have still some relevant effects in this less dense scenario.

Figure 3 shows the eigenmodal dispersion of effective constitutive parameters for this array for $\gamma = 0.45$. The top panel compares: the effective permittivity $\varepsilon_{eff}$ (solid black line); $\varepsilon_{em}$, calculated after neglecting the magnetoelectric coupling coefficient $C'_{em}$, as in Eq. (38) (dashed); $\varepsilon_{loc}$, calculated neglecting also the effects of spatial dispersion in $C_{int}$, but still considering its dependence on $\omega$ for $\beta = 0$ (dash-dotted); $\varepsilon_{CM}$, obtained using the quasi-static expression for $C_{int}$ given in (40) (dotted), which coincides with the Clausius-Mossotti definition for periodic arrays derived in [5]; finally $\varepsilon_{eq}$ (solid light green), defined in (28). All these expressions yield a purely real permittivity, as expected from the lossless assumption. However, $\varepsilon_{CM}$ rapidly diverges from the first-principle permittivity $\varepsilon_{eff}$. The value of $\varepsilon_{eff}$ actually decreases with frequency for any $k_0 d < 0.65$, due to the small noncausal feature introduced by the dipolar approximations used here, particularly relevant in the case of more densely packed arrays [42]. Magnetoelectric coupling has very little relevance here, as $\varepsilon_{em}$ practically overlaps with $\varepsilon_{eff}$, but the effects of spatial and frequency dispersion of the interaction constants are quite relevant, as seen by comparing $\varepsilon_{loc}$ and $\varepsilon_{CM}$ with $\varepsilon_{eff}$. Finally, the divergence of $\varepsilon_{eq}$ from the correct value $\varepsilon_{eff}$ is a symptom of non-negligible spatial dispersion and magnetoelectric coupling in the array, which are evidently not negligible in such dense arrays.

In comparison, the permeability is accurately predicted by all approximate models, and even the local or Clausius-Mossotti approximations predict extremely well its weak dispersion, due to the significantly lower magnetic response of the spheres all over the frequency range of interest. Interestingly, only $\mu_{eq}$ shows a moderate deviation from $\mu_{eff}$, which highlights how the effects of $\chi_{eff}^o$ may not be neglected even in this long-wavelength regime. Finally, the value of $\chi_{eff}^o$ (bottom panel) becomes relevant only towards the higher end of this frequency range, explaining the divergence of effective and equivalent parameters.

Figure 4 calculates the secondary effective parameters of this material, obtained using the different homogenization models of Fig. 3. In particular, Fig. 4a compares the exact value of normalized wave number $\beta / k_0$, as from Fig. 2a, with the approximate values $\beta_i / k_0 = \sqrt{\varepsilon_i \mu_i}$, where the pedix $i$ stands for any of the approximate models used in Fig. 3 ($i = eff, em, CM, loc$). This plot offers several interesting insights: first of all, it is noticed that $\beta_{eff}$ follows extremely well the dispersion of $\beta_{em}$, consistent with the weak effects of $C'_{em}$ on the effective parameters. However, both curves moderately diverge from the correct value $\beta / k_0$ in the range $0.5 < k_0 d < 1$, confirming that the effects of $\chi_{eff}^o$ cannot be neglected in this frequency range. The Clausius-Mossotti model $\beta_{CM}$ fails even more substantially. Fig. 4b compares the corresponding values of effective characteristic impedance $\eta_i / \eta_0 = \sqrt{\mu_i / \varepsilon_i}$. As noticed in the previous section, $\chi_{eff}^o$ does not play a direct role in the impedance when $\chi_{eff}^e = 0$, and therefore the parameters obtained neglecting $C'_{em}$ yield an accurate approximation of the effective impedance $\eta_{eff}$. It should be noted, however, that the relation between $\eta_{eff}$ and $P_{av}/M_{av}$ may not be assumed as simple as (39), due to the effects of $\chi_{eff}^o$ for relatively larger frequencies. As seen in this figure and discussed in Section 4, the equivalent parameters, despite hiding the magnetoelectric effects, predict correct values of the secondary parameters of the array, consistent with (29)-(30). This ensures that their use for scattering purposes in absence of embedded sources is perfectly legitimate, if one avoids assigning them the physical meaning of local permittivity and permeability.

In the less dense array case of Fig. 2b (not reported here for brevity), as expectable the effects of nonlocality and spatial dispersion are much less relevant, but still Clausius-Mossotti homogenization formulas would considerably deviate from the effective parameters.

Consider now the second metamaterial of interest, composed of spheres with $\varepsilon_r = 120$, which support a magnetic and an electric resonance within the low frequency range considered here. Figure 5 shows the eigenwave number dispersion for such array with $\gamma = 0.45$, with symbols analogous to Fig. 2. It is immediately recognized that the exact dispersion of normalized wave number $\beta/k_0$ (solid lines) is much more intricate than in the previous example. As expected, $\beta/k_0$ initially grows with frequency, until hitting the first band-gap of the array at $(k_0 d) = 0.594$, at the magnetic resonance frequency $\alpha_m^{-1} - C = 0$. The narrow frequency region within the bandgap should be completely disregarded in terms of homogenization, since, as discussed above, the effects of array granularity plays a major role here. Passed the magnetic bandgap, a branch with imaginary wave number $\beta = i\beta_i$ is entered (thin solid line), which connects with the next real branch at $(k_0 d) = 0.723$, at the point for which $\beta = 0$. The following bandgap is then hit at the electric resonance frequency $\alpha_e^{-1} - C = 0$, at $(k_0 d) = 0.891$, and the next real branch is obtained at $(k_0 d) = 0.909$. As seen in Fig. 5, this behavior is well described by approximate dispersion relations, after neglecting the effects of $C'_{em}$ or the spatial dispersion in $C_{int}$, as in $\beta_{em}$ and $\beta_{CM}$ respectively. This is due to the fact that in this array the local inclusion resonances dominate the array response and hide weak spatial dispersion effects at the lattice level. Of course, in this scenario it is not possible to neglect the magnetic effects in the dielectric particles, as for $\beta_e$ and $\beta_{e-CM}$, since this would completely miss the first magnetic bandgap resonance.

The effective constitutive parameters of this array are shown in Fig. 6, with analogous symbols as described in Fig. 3. Even if the spatial dispersion effects are negligible in evaluating $\beta(\omega)$ in Fig. 5, they play a major role in the correct definition of constitutive parameters, in particular near the electric and magnetic resonances of the inclusions. First, it is noticed that Clausius-Mossotti formulas completely miss the relevant magnetoelectric coupling arising near the bandgaps, and the permittivity especially suffers of this approximation, starting from very low frequencies. Towards the first (magnetic) resonance, $\varepsilon_{em}$ may approximate relatively well the effective permittivity $\varepsilon_{eff}$, confirming that the effect of $C'_{em}$ is small on the permittivity dispersion, dominated by the local inclusion resonances. However, the value of $\chi_{eff}^{o}$ assumes large values near the two resonances and it cannot be neglected. Near the magnetic resonance, the effective permittivity experiences a sharp Lorentzian resonance, completely missed by $\varepsilon_{CM}$ and even by $\varepsilon_{loc}$, which is an evident symptom of magnetoelectric coupling in the array. In contrast, the various models for magnetic permeability all have good agreements with the effective model (with the exception of a small resonant feature arising at the electric bandgap resonance of the array). In the region where $\beta$ is imaginary, immediately following the bandgaps, all the models correctly predict a negative effective permeability or permittivity region, which crosses zero at $(k_0 d) = 0.723$ and $(k_0 d) = 0.909$, together with the value of $\beta$. In this negative parameter range, as expected, $\chi_{eff}^{o}$ is imaginary (dashed lines in the bottom panel), which ensures that the equivalent parameters are real quantities (one of them negative).

Special attention should be paid to the dispersion of the equivalent permittivity $\varepsilon_{eq}$ in Fig. 6a (lighter green line). Its slope is negative all the way until the magnetic bandgap, producing an anti-resonant dispersion consistent with usual artifacts arising in common retrieval procedures near magnetic resonances [20]-[29]. It is evident that these effects are associated with $\chi_{eff}^{o}$, hidden in the definition of equivalent permittivity. It is true that the equivalent parameters may describe well the secondary parameters of the array, but their physical meaning in this case considerably diverge from the first-principle definition of permittivity and permeability. A simple homogenization model based on the equivalent representation would fail to capture the physics of the array near the bandgap resonance, predicting $\varepsilon_{eq} = \varepsilon_0$ (43), when in reality the averaged polarization vector has a strong resonance. These effects are discussed in more detail in [32].

If the discrepancy between $\varepsilon_{eq}$ and $\varepsilon_{eff}$ was expected near resonance, another transition region in which the equivalent parameters $\varepsilon_{eq}, \mu_{eq}$ lose their physical meaning is the region near $(k_0 d) = 0.723$, for which $\beta \simeq 0$. As confirmed by Fig. 6a, and consistent with the analysis in Section 6a, in this region

$$\varepsilon_{eff} \simeq \varepsilon_{em} \simeq \varepsilon_{loc}$$
$$\mu_{eff} \simeq 0 \qquad . \qquad (47)$$
$$\chi^o_{eff} \simeq 0$$

Indeed, the correct value of effective permittivity coincides with the local value $\varepsilon_{loc}$, since $\beta \simeq 0$, but this value is substantially different from $\varepsilon_{eq}$. This is due to the fact that, although the magnetoelectric coefficient is near zero, the ratio $\chi^o_{eff} / \beta = \kappa_{eff}$ is finite, causing $\varepsilon_{eq}$ to diverge from $\varepsilon_{eff}$ and to lose its meaning of average electric polarizability. In this near-zero index region, the weak spatial dispersion captured by $\kappa_{eff}$ in (36) cannot be neglected, even if the effective wavelength is very large. This confirms the results in [26] derived for periodic arrays of split-ring resonators, in which the presence of non-negligible spatial dispersion effects in the long-wavelength ($\beta d \simeq 0$) scenario is discussed.

Figure 7 shows the dispersion of the effective index of refraction and characteristic impedance obtained through the various parameters of Fig. 6, similar to Fig. 4. All the curves agree with high accuracy within the real branches, since their dispersion is dominated by the local resonances at the inclusion level. This example clearly shows that indeed $\beta$ and $\eta$ for this array may be easily derived applying local concepts, like Clausius-Mossotti relations or simple retrieval procedures, since they are dominated by local resonances at the inclusion level; however, inferring from these secondary parameters the physical values of permittivity and permeability, as commonly done in standard homogenization techniques, leads to physical artifacts and inconsistencies [32].

As a third example, we consider the case of an array of conducting particles. Figure 8 shows the dispersion of wave numbers for $\gamma = 0.45$ (a) and $\gamma = 0.3$ (b), analogous to Fig. 2. In this case, the wave numbers predicted using just electric effects of the particles are evidently incorrect, since the magnetic contribution for conducting particles is never negligible. Moreover, the effect of the coupling coefficient $C'_{em}$ is particularly relevant in this conducting scenario, which shows significant divergence between $\beta$ and $\beta_{em}$, due to the relevance of the magnetic effects even at very low frequencies.

Figure 9 shows the corresponding constitutive parameters for the case $\gamma = 0.45$. $\varepsilon_{eff}$ also in this scenario shows a distinctly negative slope, all over the range of frequencies considered here, due to small noncausal features introduced by the polarizability model [42]. This is compensated by the positive slope of the effective permeability, which assumes, as expected, diamagnetic values [57]. Only the Clausius-Mossotti quasi-static model $\varepsilon_{CM}$ predicts a permittivity with positive slope, whereas all the other models consistently follow the trend of $\varepsilon_{eff}$. To confirm the strong influence of $\chi^o_{eff}$, the equivalent parameters $\varepsilon_{eq}$ and particularly $\mu_{eq}$ considerably deviate from the effective parameters. Figure 10, finally, shows the dispersion of the calculated wave numbers and characteristic impedances obtained using the effective constitutive parameters of Fig. 9. It is seen how all curves agree reasonably well with the exact dispersion of $\eta_{eff}$, except the quasi-static Clausius-Mossotti formula, which neglects frequency and spatial dispersion effects of the interaction constants. The divergence of all the curves from the exact dispersion of $\beta$ is particularly striking, as a symptom of the relevance of the magnetoelectric coefficient $\chi_{eff}^{o}$ in this example. We have also analyzed the less dense configuration $\gamma = 0.3$, as in Fig. 8b (not reported here for brevity), which indeed provides analogous results, but less strong variations from the background parameters, similar to the previous examples.

## 8. Conclusions

We have laid out here from first-principles a general homogenization theory to define the effective constitutive parameters of periodic metamaterials. Our theory can describe periodic arrays of arbitrary inclusions within a homogenized model that has been proven not to depend on the external form of excitation and to preserve the physical meaning of constitutive parameters, overcoming and correcting several limitations and artifacts of other homogenization approaches. The present theory effectively combines the rigorous approach of Floquet-based homogenization theories with the advantages of locality and general applicability of less accurate retrieval techniques. We have distinguished between a rigorous and general description of metamaterials, based on their *effective* constitutive parameters, which inherently require taking into account weak spatial dispersion effects in the form of magnetoelectric coupling at the lattice level, and a simpler *equivalent* constitutive model, applicable only to eigenmodal propagation and consistent with standard retrieval techniques. Our theory shows that the commonly used *equivalent* representation can accurately capture the secondary parameters of the array, implying that in absence of embedded sources they can provide a reasonable description of the array for scattering purposes. However, they should not be used to deduce the permittivity and permeability of the array, as their physical meaning is severely limited by the presence of hidden spatial dispersion effects [32], which have been revealed here. A rigorous retrieval procedure to extract the first-principle *effective* parameters from scattering measurements will be presented in an upcoming paper. Although our theory is very general, the numerical results presented here have focused on isotropic arrays, center-symmetric inclusions, lossless dielectric materials and eigenmodal propagation within a dipolar model, in order to better highlight the specific effects of spatial dispersion, neglected in simpler homogenization models. For space constraints, we have not discussed here the effects of losses, of magnetodielectric and bianisotropic inclusions, of impressed sources, of non-TEM propagation and of higher-order multipoles, which will be analyzed separately. We have applied the present theory to model finite metamaterial devices in [58]-[59].

**Acknowledgments**

We thank Arthur D. Yaghjian, Robert A. Shore, Mario G. Silveirinha and Mayer Landau for relevant and fruitful discussions. This work has been supported by the U.S. AFOSR YIP award No. FA9550-11-1-0009, the U.S. Air Force Research Laboratory with contract No. FA8718-09-C-0061, the NSF CAREER award No. ECCS-0953311 and by the ONR MURI grant No. N00014-10-1-0942.

[40] It may be proven that this is the case for isotropic unit cells and lattices, as we assume in the following, when $\hat{\boldsymbol{\beta}}$ is along one of the lattice axes and, more generally, for any propagation direction in the limit $k_0 d \ll 1$.

[41] The reader should be aware at this point that the introduction of polarizability models to approximate the inclusions and their interactions inherently introduces a small noncausality in the metamaterial response, whose causes are explained in detail in [42]. This artifact is usually negligible, but it becomes relevant in the case of densely packed arrays, as in some of the numerical examples of Section 7.

[54] In the case of degenerate resonances $\alpha_e^{-1} = \alpha_m^{-1} = C(\omega, \pi/d)$, some of the following derivations in this section do not apply. This special circumstance, which can arise in the case of impedance-matched inclusions, will be considered in a separate contribution.

[55] For an example involving magnetodielectric spheres, the interested reader can refer to [32].

[56] A. Alù, and N. Engheta, *J Appl Phys*. **97**, 094310 (2005).

[57] E. N. Economou, Th. Koschny, and C. M. Soukoulis, *Phys. Rev. B* **77**, 092401 (2008).

[58] X X. Liu, and A. Alù, *J. Nanophoton.*, in press (2011).

[59] X. X. Liu, and A. Alù, *Metamaterials*, in press (2011).

**Figures**

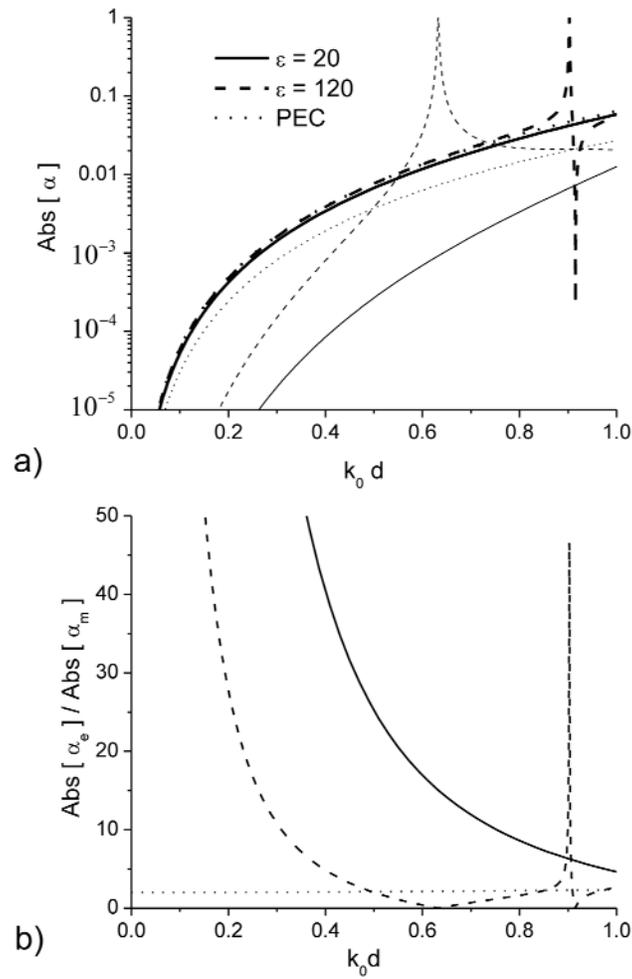

**Figure 1** – (a) Frequency dispersion of the electric (thick line) and magnetic (thin) normalized polarizability of the individual inclusions for the metamaterial arrays considered in the following figures: (solid) dielectric spheres with permittivity $\varepsilon = 20\varepsilon_0$; (dashed) dielectric spheres with $\varepsilon = 120\varepsilon_0$; (dotted) conducting spheres; (b) Ratio of electric over magnetic polarizability for the same geometries. Here $\gamma = 0.45$.

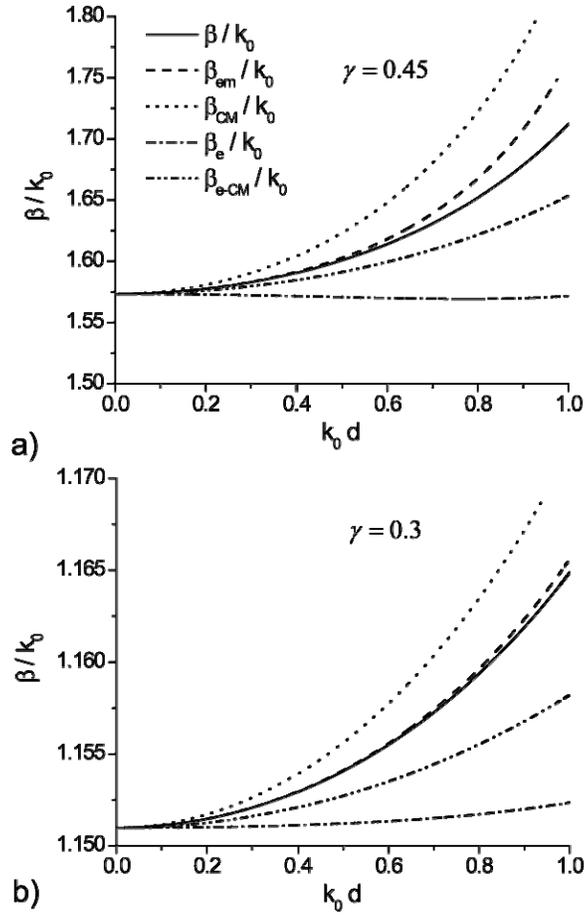

**Figure 2** – Frequency dispersion of the guided wave number, and its approximations as defined in the text, for an array of dielectric spheres with $\varepsilon = 20\varepsilon_0$, with (a) $\gamma = 0.45$, (b) $\gamma = 0.3$.

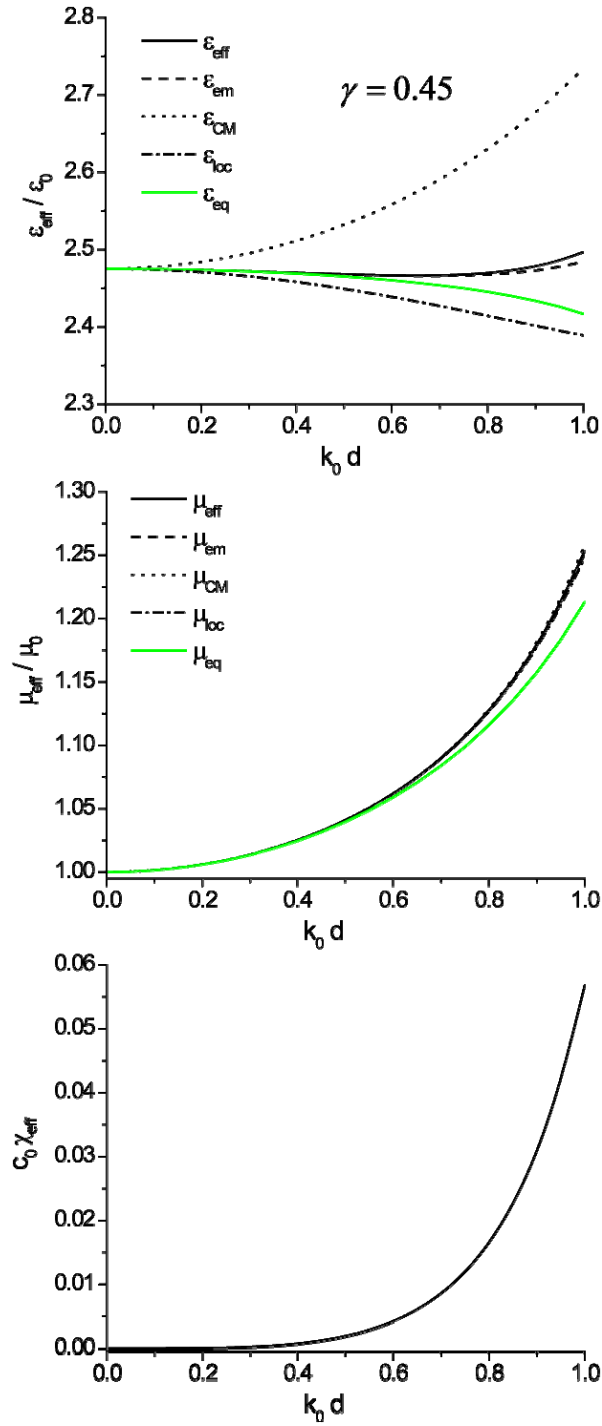

**Figure 3** – (Color online): Frequency dispersion of the effective constitutive parameters, and their approximations as defined in the text, for the array of Fig. 2a.

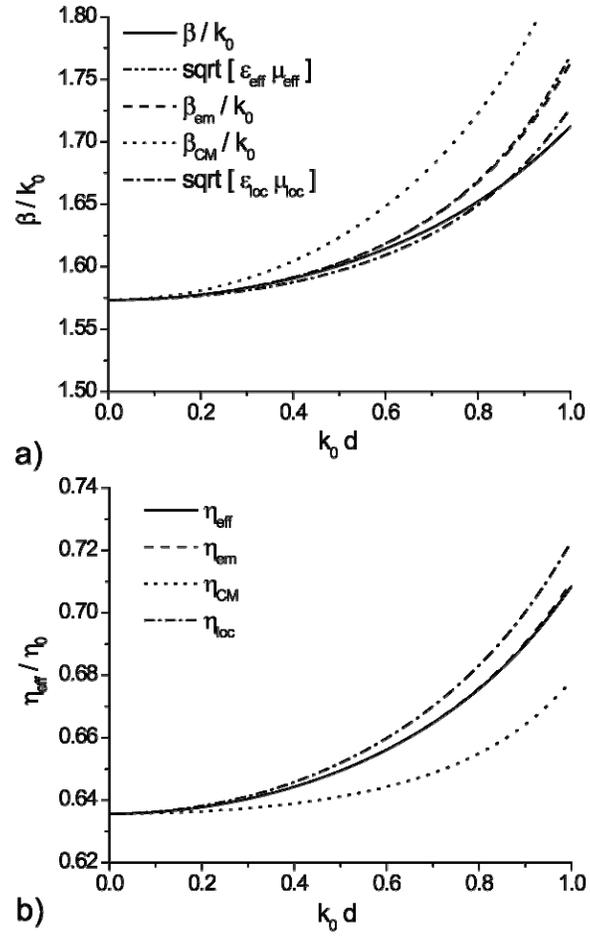

**Figure 4** – Frequency dispersion of the effective wave number and characteristic impedance calculated from the constitutive parameters of Fig. 3.

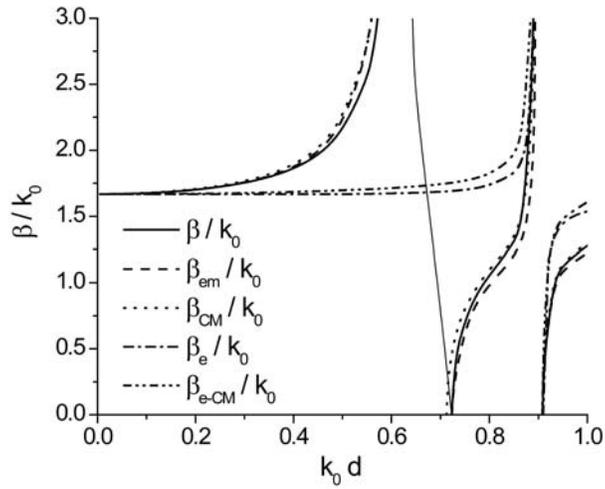

**Figure 5** – Frequency dispersion of the guided wave number, and its approximations as defined in the text, for an array of dielectric spheres with $\varepsilon = 120\varepsilon_0$ and $\gamma = 0.45$. The thin solid line corresponds to the imaginary branch of $\beta$.

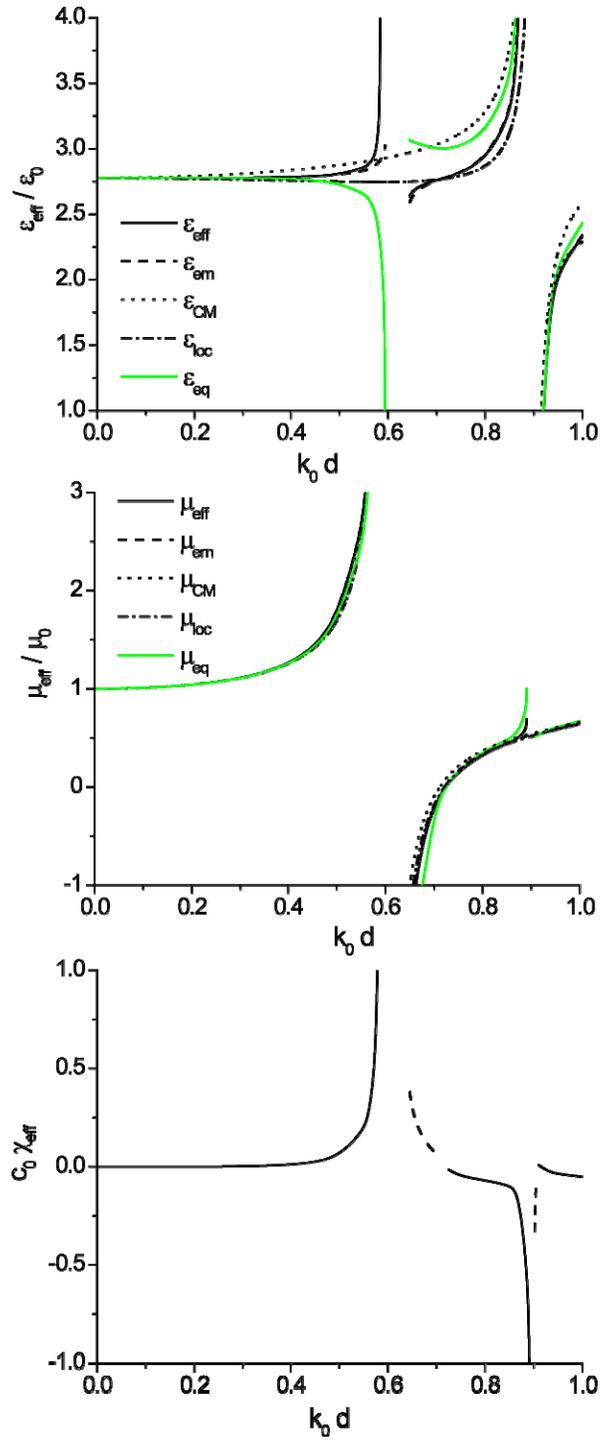

**Figure 6** – (Color online): Frequency dispersion of the effective constitutive parameters, and their approximations as defined in the text, for the array of Fig. 5. Dashed lines in the bottom panel refer to branches with imaginary values.

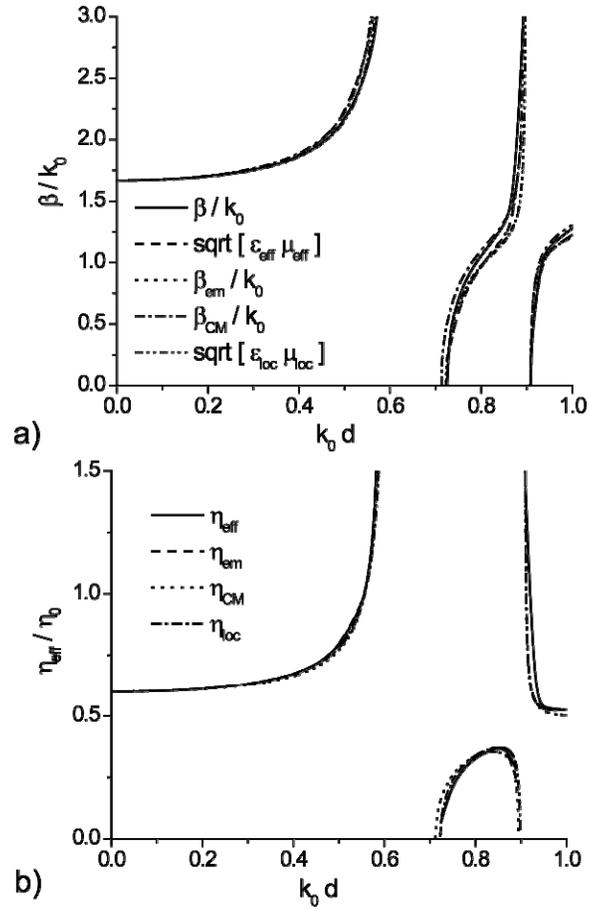

**Figure 7** – (Color online): Frequency dispersion of the effective wave number and characteristic impedance calculated from the constitutive parameters of Fig. 6.

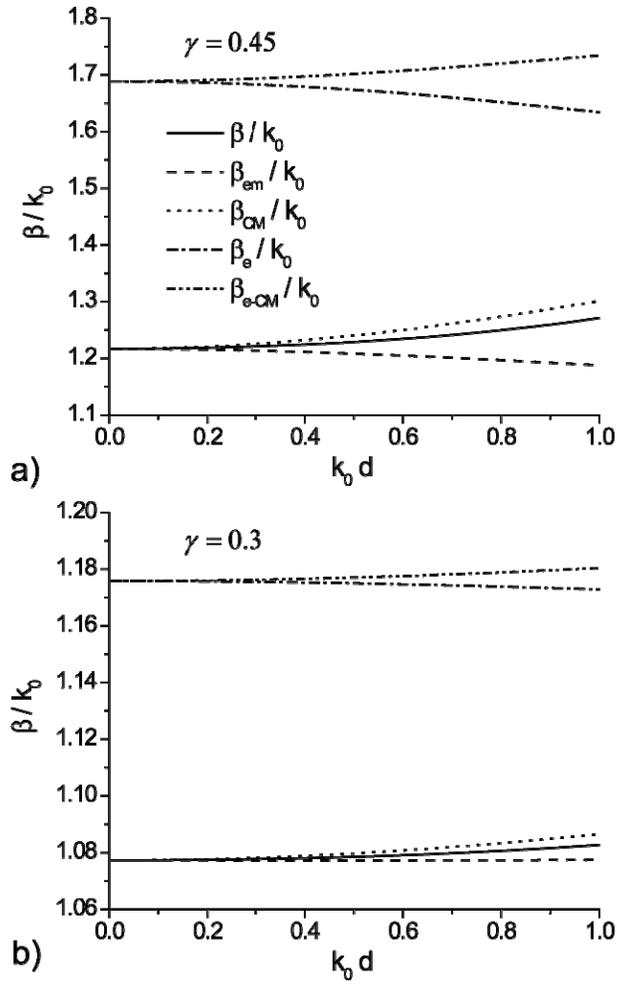

**Figure 8** – Frequency dispersion of the guided wave number, and its approximations as defined in the text, with frequency for an array of conducting spheres, for (a) $\gamma = 0.45$, (b) $\gamma = 0.3$.

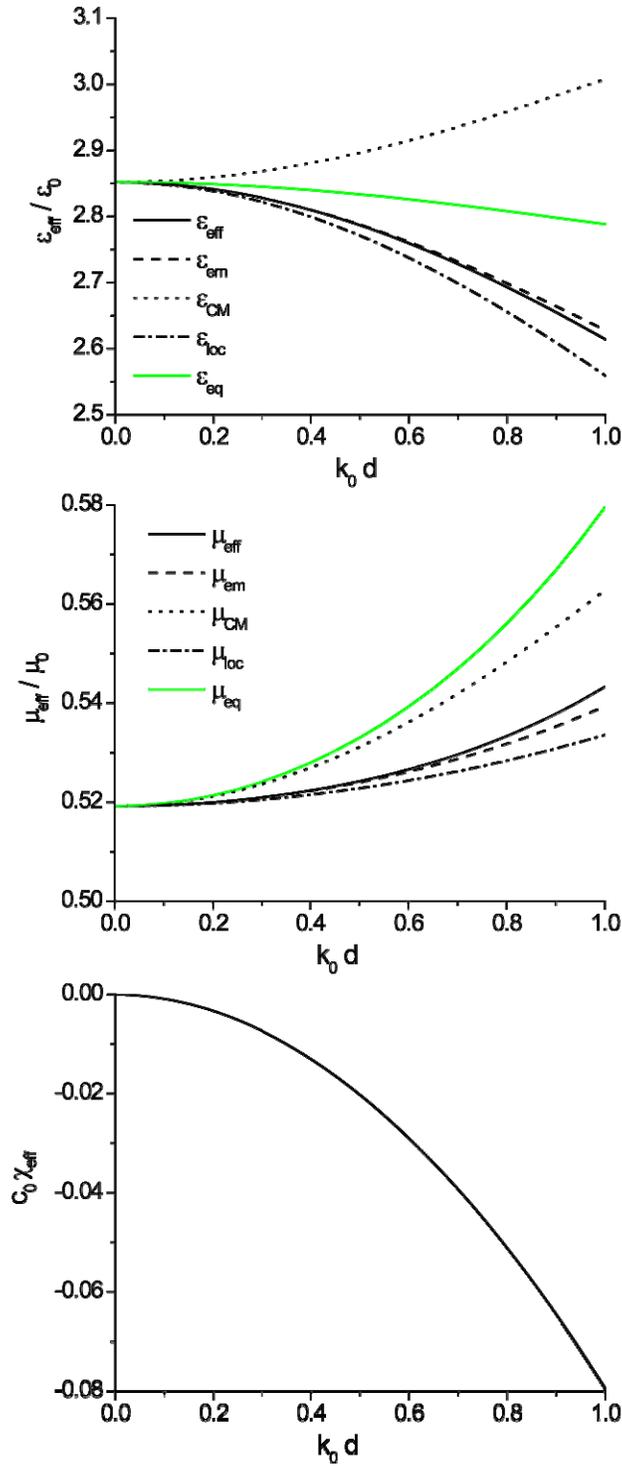

**Figure 9** – (Color online): Frequency dispersion of the effective constitutive parameters, and their approximations as defined in the text, for the array of Fig. 8a.

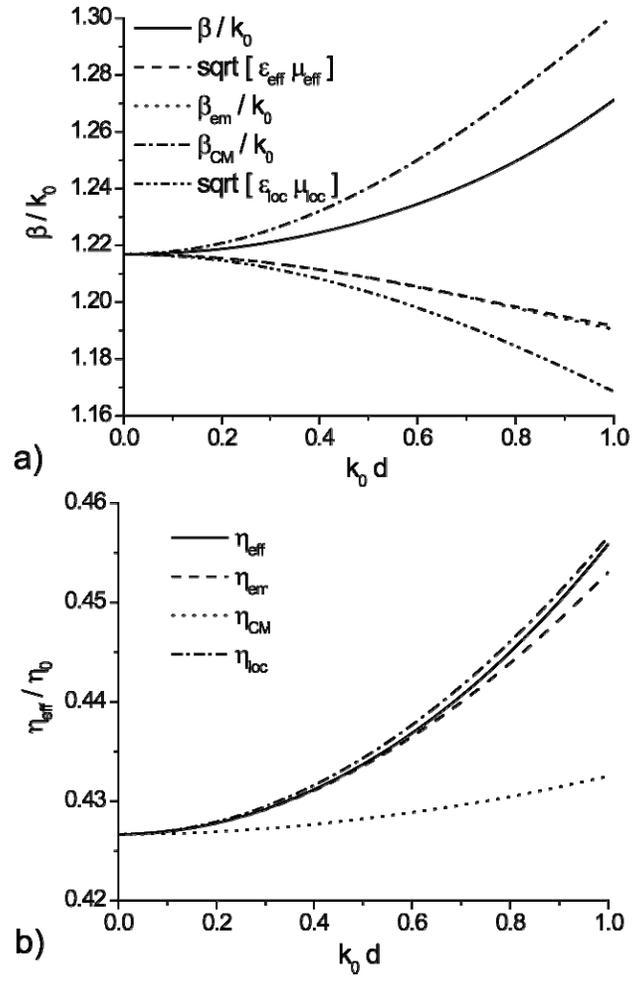

**Figure 10** – Frequency dispersion of the effective wave number and characteristic impedance calculated from the constitutive parameters of Fig. 9.